%% file: template.tex
\documentclass[journal]{vgtc}                  % final (journal style)

\ifpdf%                                % if we use pdflatex
  \pdfoutput=1\relax                   % create PDFs from pdfLaTeX
  \usepackage{graphicx}                % allow us to embed graphics files
  \DeclareGraphicsExtensions{.pdf,.png,.jpg,.jpeg} % for pdflatex we expect .pdf, .png, or .jpg files
\else%                                 % else we use pure latex
  \usepackage{graphicx}                % allow us to embed graphics files
  \DeclareGraphicsExtensions{.eps}     % for pure latex we expect eps files
\fi%

\graphicspath{{figures/}{pictures/}{images/}{./}} % where to search for the images
\usepackage{microtype}                 % use micro-typography (slightly more compact, better to read)
\PassOptionsToPackage{warn}{textcomp}  % to address font issues with \textrightarrow
\usepackage{textcomp}                  % use better special symbols
\usepackage{mathptmx}                  % use matching math font
\usepackage{times}                     % we use Times as the main font
\usepackage{cite}                      % needed to automatically sort the references
\usepackage{mathrsfs}
\usepackage[normalem]{ulem}
\usepackage{xspace}
\usepackage{lipsum}
\usepackage{amsmath}
\usepackage{enumitem}
\usepackage{algorithm}
\usepackage{algorithmicx}
\usepackage[noend]{algpseudocode}
\usepackage{xspace}
\usepackage{graphicx}
\usepackage{duckuments}
\usepackage{multirow}
\usepackage{wrapfig}
\newcommand{\zq}[1]{{\color{black}{#1}}}
\onlineid{1150}

\vgtccategory{Research}

\vgtcpapertype{Representations \& Interaction}

\title{CompositingVis: Exploring Interactions for \\ Creating Composite Visualizations in Immersive Environments}
\author{%
  \authororcid{Qian Zhu}{0000-0001-5108-3414},
   \authororcid{Tao Lu}{0009-0001-3078-5975}, \authororcid{Shunan Guo}{0000-0001-5355-8399}, \authororcid{Xiaojuan Ma}{0000-0002-9847-7784}, \authororcid{Yalong Yang}{0000-0001-9414-9911}
}

\authorfooter{
  \item Qian Zhu and Xiaojuan Ma are with the Hong Kong University of Science and Technology.
        E-mail: qian.zhu@connect.ust.hk and mxj@cse.ust.hk.
  \item Tao Lu and Yalong Yang are with Georgia Institute of Technology.
  	E-mail: \{luttul, yalong.yang\}@gatech.edu.
  \item Shunan Guo is with Adobe Research.
  	E-mail: sguo@adobe.com
}

\abstract{
Composite visualization represents a widely embraced design that combines multiple visual representations to create an integrated view. 
However, the traditional approach of creating composite visualizations in immersive environments typically occurs asynchronously outside of the immersive space and is carried out by experienced experts. 
In this work, we aim to empower users to participate in the creation of composite visualization within immersive environments through embodied interactions. 
This could provide a flexible and fluid experience with immersive visualization and has the potential to facilitate understanding of the relationship between visualization views.
We begin with developing a design space of embodied interactions to create various types of composite visualizations with the consideration of data relationships. Drawing inspiration from people's natural experience of manipulating physical objects, we design interactions \zq{based on the combination of 3D manipulations} in immersive environments.
Building upon the design space, we present a series of case studies showcasing the interaction to create different kinds of composite visualizations in virtual reality. 
Subsequently, we conduct a user study to evaluate the usability of the derived interaction techniques and user experience of creating composite visualizations through embodied interactions. 
We find that empowering users to participate in composite visualizations through embodied interactions enables them to flexibly leverage different visualization views for understanding and communicating the relationships between different views, which underscores the potential of several future application scenarios.
}


\teaser{
  \centering
  \includegraphics[width=\linewidth]{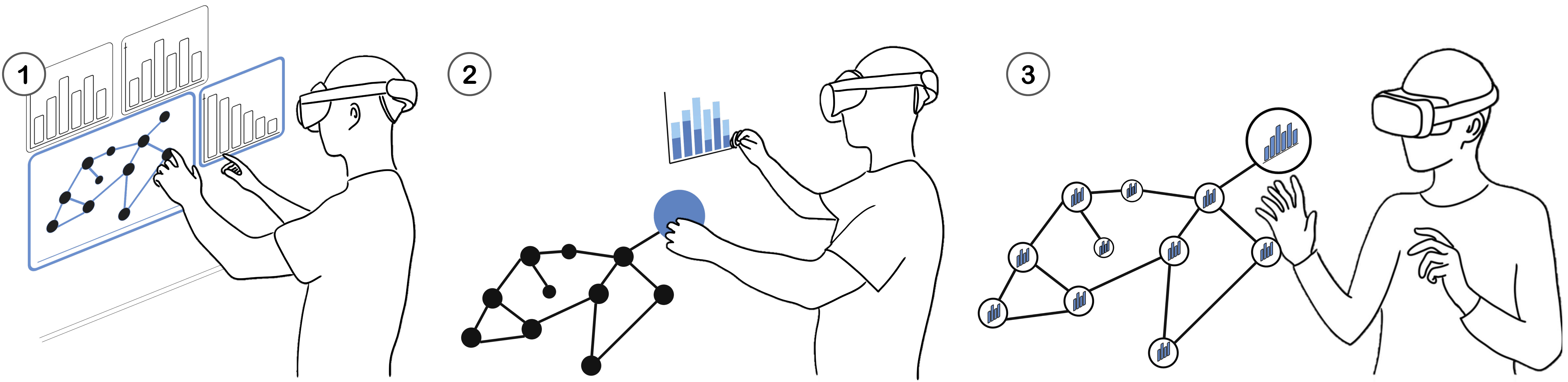}
  \caption{
  A conceptual scenario of creating a composite visualization in virtual or augmented reality. (1) A user is viewing multiple visualizations, including a graph visualization and bar charts, representing data with some underlying association. (2) The user wants to create a composite visualization that includes both views for analysis. The user grabs the views of interest and assembles them like compositing entities. (3) After composition, a composite visualization appears, integrating both graphical and bar chart visualizations.
  }
  \label{fig:teaser}
  \vspace{0.8em}
}

\graphicspath{{figs/}{figures/}{pictures/}{images/}{./}} 
\usepackage{tabu}                      % only used for the table example
\usepackage{booktabs}                  % only used for the table example
\usepackage{lipsum}                    % used to generate placeholder text
\usepackage{mwe}                       % used to generate placeholder figures

\usepackage{mathptmx}                  % use matching math font
\usepackage{enumitem}
\setitemize{noitemsep,topsep=0pt,parsep=0pt,partopsep=0pt}

\begin{document}

%%%%%%%%%%%%%%%%%%%%%%%%%%%%%%%%%%%%%%%%%%%%%%%%%%%%%%%%%%%%%%%%
%%%%%%%%%%%%%%%%%%%%%% START OF THE PAPER %%%%%%%%%%%%%%%%%%%%%%
%%%%%%%%%%%%%%%%%%%%%%%%%%%%%%%%%%%%%%%%%%%%%%%%%%%%%%%%%%%%%%%%

\maketitle

\input{sections/1_Introduction.tex}
\input{sections/2_RelatedWork.tex}
\input{sections/3_Concepts.tex}

\input{sections/4_DesignSpace}

\input{sections/5_Dev}
\input{sections/6_Study}

\input{sections/7_Discussion.tex}
\input{sections/8_Conclusion.tex}
\input{sections/9_Ack.tex}

\bibliographystyle{abbrv-doi-hyperref}
% \bibliographystyle{unsrtnat}  
% \bibliographystyle{abbrv-doi-hyperref-narrow}
% %\bibliographystyle{abbrv-doi-narrow}
% \bibliographystyle{plainnat}
% \bibliographystyle{abbrv-doi}
\bibliography{template}

% \appendix % You can use the `hideappendix` class option to skip everything after \appendix

% \section{About Appendices}
% Refer to \cref{sec:appendices_inst} for instructions regarding appendices.

% \section{Troubleshooting}
% \label{appendix:troubleshooting}

% \subsection{ifpdf error}

% If you receive compilation errors along the lines of \texttt{Package ifpdf Error: Name clash, \textbackslash ifpdf is already defined} then please add a new line \verb|\let\ifpdf\relax| right after the \verb|\documentclass[journal]{vgtc}| call.
% Note that your error is due to packages you use that define \verb|\ifpdf| which is obsolete (the result is that \verb|\ifpdf| is defined twice); these packages should be changed to use \verb|ifpdf| package instead.

% \subsection{\texttt{pdfendlink} error}

% Occasionally (for some \LaTeX\ distributions) this hyper-linked bib\TeX\ style may lead to \textbf{compilation errors} (\texttt{pdfendlink ended up in different nesting level ...}) if a reference entry is broken across two pages (due to a bug in \verb|hyperref|).
% In this case, make sure you have the latest version of the \verb|hyperref| package (i.e.\ update your \LaTeX\ installation/packages) or, alternatively, revert back to \verb|\bibliographystyle{abbrv-doi}| (at the expense of removing hyperlinks from the bibliography) and try \verb|\bibliographystyle{abbrv-doi-hyperref}| again after some more editing.

\end{document}

%% file: sections/1_Introduction.tex
\section{Introduction} \label{sec:introduction}
As the volume and complexity of data continue to grow, the demand for sophisticated data visualization has escalated to tackle complex analytical tasks. 
This often requires the integration of multiple visual representations, which facilitates a comprehensive understanding of the relationships between different data facets and visualization views. 
Consequently, significant research efforts have been devoted to combining multiple visual representations to form a coherent and meaningful layout~\cite{deng2022revisiting}. This extensively embraced design strategy is commonly referred to as composite visualization~\cite{javed2012exploring, elmqvist2011embodied}.
% This widely adopted design strategy integrates visualizations by capitalizing on their strengths and mitigating their limitations, commonly known as composite visualization~\cite{elmqvist2011embodied}.
% For example, medical researchers frequently encounter complex datasets comprising patient demographics, treatment outcomes, and disease prevalence trends. 

Composite visualization has also been preliminarily explored in immersive environments, with the rapid development of Immersive Analysis (IA)~\cite{marriott2018immersive, saffo2023unraveling,zhao2022metaverse,ens2021grand}.
% As immersive analytics (IA)~\cite{marriott2018immersive, saffo2023unraveling} continue to evolve rapidly, analysts are encountering complex tasks in immersive environments and consequently requiring composite visualization solutions. 
While previous research demonstrates the advantages of composite visualizations for IA, the workflow of creating such visualizations relies on pre-construction on computers by visualization designers with coding expertise~\cite{yang2020tilt, yang2020embodied}.
This approach usually results in a passive user experience, where users interact with composite visualizations that were created asynchronously, rather than engaging in the creation process\cite{sicat2018dxr, cordeil2019iatk, butcher2020vria}.
As it is difficult to anticipate visual representation needs and their combinations during the Exploratory Data Analysis (EDA) stage in IA, it is crucial to empower users with the flexibility to integrate various primitive visualizations into composite views. This flexibility enables users to freely explore, formulate hypotheses, and validate their ideas, facilitating an engaging and fluid experience with data visualization~\cite{elmqvist2011fluid}. 
% Update3.18 why let them create? 1. engagement, 2. interactive UX for data understanding and memory, 3. flexibility to explore data, formulate hypotheses, and conduct experiments to discover patterns and relationships hidden behind the data.

Immersive environments present a unique opportunity to facilitate a fluid experience with composite visualizations, as they offer large display spaces and embodied interaction~\cite{marriottCh2immersive, marriott2018immersive}. 
Capitalizing on these advantages, our work aims to engage users in the creation of composite visualizations in immersive environments by employing two design metaphors. 
First, we take composite visualizations as constructs based on the composition of multiple primitive views (e.g., bar charts or scatterplots). This perspective also inherently allows for the deconstruction of a composite visualization into its constituent primitive visualizations~\cite{javed2012exploring}.
Second, we propose empowering users with a \textit{``superpower''}~\cite{willett2021perception} to construct a composite view with primitive views. This process is akin to the natural interactions used in the assembly or piecing together of physical objects, enhancing the intuitiveness and engagement of the composition process~\cite{endert2012unifying}.
However, designing proper interactions for the composition of visualizations is challenging. 
One consideration is that determining whether different visualization views can be combined into a composite view is influenced not merely by the type of composite view but also by the inherent constraints of the underlying data relationships~\cite{yang2014understand}.
In addition, different types of composite visualizations represent distinct data relationships (e.g., nested views and juxtaposed views can encode different relationships~\cite{javed2012exploring}), demanding the development of specialized interaction techniques tailored to each type. Moreover, the interactions need to prevent ambiguity and accurately convey user intentions of compositing visualization views, while ensuring an intuitive and fluid user experience~\cite{elmqvist2011fluid}.

To bridge this gap, we prioritize the data relationships and connect them to the creation of composite visualizations. We then draw on the natural human skill of spatially manipulating physical objects as a metaphor to develop the design space of interactions for \zq{authoring} composite visualizations. 
To demonstrate the design space, we implement a set of proof-of-concept cases with various types of composite visualization in Virtual Reality (VR). Finally, we conduct a user study to 1) assess the usability of the derived interactions, and 2) evaluate user experience to explore the potential advantages and problems of involving users in the embodied composition process. 
Our goal is to provide insights and design recommendations for developers or designers in crafting interactive experiences with composite visualization in IA systems.
The main contributions of this paper are:
\begin{itemize}
    \item Development of a design space that considers embodied interactions and data relationships to create composite visualization in immersive environments;
    \item Implementation of a set of cases with usage scenarios in VR, which demonstrates the utility of the design space;
    \item A user study that assesses the usability of representative embodied interactions derived from our design space and offers guidelines for crafting interactive experience with composite visualizations. 
\end{itemize}

%% file: sections/2_RelatedWork.tex
\section{Related Work} \label{sec:related-work}
\textbf{Composite Visualization in Immersive Environments.}
Javed and Elmqvist introduced the composite visualization concept with the categorization for specifying the spatial compositions of multiple visualizations in the same visual space~\cite{javed2012exploring}.
For example, one widely kind of composite visualization is juxtaposed views, which place multiple visualization views side-by-side~\cite{roberts2007state, baldonado2000guidelines}.
Over the past decade, significant research efforts have been directed to design the visual representations of composite visualizations that integrate multiple data visualization views to present multi-faceted data~\cite{deng2022revisiting}. 
Composite visualization gained considerable attention due to its capacity to leverage the strengths of different views or integrate multiple views to mitigate their respective weaknesses~\cite{javed2012exploring, yang2014understand}.

In recent years, researchers in the emerging research field of Immersive Analytics (IA) have begun to explore how to design or display composite visualizations within immersive spaces~\cite{yang2020tilt, cordeil2017imaxes, liu2020design, hubenschmid2021stream, prouzeau2019visual, langner2021marvis}.
For instance, Liu et al. investigated the design of \textit{juxtaposed} views represented by small multiples in immersive environments~\cite{liu2023datadancing, liu2020design, liu2022effects}. Hubenschmid et al. explored interaction techniques for the \textit{integrated} views with explicit visual links in immersive spaces~\cite{hubenschmid2021stream}. 
\zq{Langner et al. combined mobile devices and augmented reality for visual data analysis with composite views, such as juxtaposed views or overloaded views~\cite{langner2021marvis}.}
Yang et al. proposed \textit{TiltMap}, a composite visualization that combines a map and a bar chart in immersive environments to efficiently present area-linked data in a \textit{superimposed} way~\cite{yang2020tilt}. 

These works revealed the benefits of immersive environments in the presentation and interaction with composite visualizations. 
However, due to the workload and coding expertise required for constructing composite visualizations, users often find it difficult to participate in the creation process. 
Yet, recognizing the critical role of user involvement in tailoring visualizations to their specific needs and preferences~\cite{kujala2003user, chen2019marvist, satyanarayan2019critical}, IA systems need to consider offering users the opportunity to actively participate in the creation of composite visualizations. We \zq{aim to} empower users in creating immersive composite visualizations from the perspective of embodied interactions.
% should be designed with intuitive, natural interactions that empower users to 
% the construction of compostie visualizations in IA is still chanllenging as it requires code expertise and usually created by developers or visualization experts.
% they only allow users to experience the created composite visualizations, rather than involving them in the process of creation,    
% which is beneficial to users' understanding and analysis of composite visualization. 
% We aim to explore how to use direct and natural embodied interactions to construct composite visualizations.

\textbf{Immersive Visualization Authoring.}
Authoring visualizations is an essential part of data analysis and communication. 
However, combining two or more views to form a composite visualization in immersive environments is a nontrivial task, even for experts.
% Previous research has mostly approached it from the perspective of creating visualizations by offering programming frameworks~\cite{wickham2010layered, satyanarayan2016vega} or commercially available visualization systems~\cite{}. 
% A full range of toolkits or methods have been developed to assist users in the creation of visualizations. 
% One type of research starts from the fundamentals based on Grammar of Graphics to assist visualization experts in flexibly writing codes to create visualizations, such as ggplot~\cite{wickham2010layered} and Vega-Lite\cite{}. 
In the field of IA, previous efforts have aimed to provide toolkits for developers or experts to author immersive visualizations~\cite{cordeil2019iatk, sicat2018dxr, butcher2020vria, fleck2022ragrug}.
For example, IATK~\cite{cordeil2019iatk} provides a toolkit that allows users to author immersive visualizations and analyze data with embodied interactions (e.g., filter). 
DXR~\cite{sicat2018dxr} is another toolkit based on Unity, which helps create immersive visualizations with declarative JSON specification. 
% However, it requires users to use Unity as the development platform. 
VRIA~\cite{butcher2020vria} offers a web-based framework building upon WebVR for creating IA experiences. 
Nevertheless, these efforts \zq{on authoring immersive visualization is often separated from the immersive environments where the visualization is actually applied} and thereby detaching the user experience between creation and analysis.
% these efforts predominantly rely on code-based approaches, which require coding expertise 
% thereby detaching the creation of immersive visualizations from users' experiences, understanding, and data analysis within immersive environments. 

Only a few works provided users with a seamless experience of creating immersive visualizations~\cite{batch2023wizualization, cheng2023ncarvis, cordeil2017imaxes}.
\zq{Satkowski et al. proposed an extended model for authoring visualizations to facilitate seamless integration of visualization creation and presentation~\cite{satkowski2021visualization}.}
% require users having coding expertise and spending time on working with the desktop to learn how to create immersive visualizations.
Cordeil et al. introduced \textit{ImAxes}~\cite{cordeil2017imaxes}, which allows users to construct multivariate data visualizations through embodied interactions. However, ImAxes is limited to constructing multidimensional data visualization based solely on axes.
\zq{Satkowski et al. explored the combination of mobile devices and AR HMDs for in-situ authoring of visualizations in an early prototype~\cite{satkowski2021towards}, enabling the configuration of visualization directly in real-world environments.}
The latest work, \textit{Wizualization}~\cite{batch2023wizualization}, leveraged magic as the metaphor to design gestures and speech interactions for authoring and analyzing immersive visualization.
However, \zq{these works} mainly focused on authoring primitive visualization views (e.g., scatterplots) rather than the embodied creation of composite visualizations. 
We aim to investigate intuitive and effective interactions that enable users to author composite visualizations in immersive environments.
% we allow a 'hands-on' understanding of composite visualization through the lens of interaction
 % D3 vega: creating vis by writing programs by professional data vis designers/experts; not for ordinary users...
%  commercial-grade vis systems like Tableau, PowerBI, charticulartor Lyra... without programming skills
%   \item {motivation of creating composite vis derived from the auto. design generation field}
%   \item {low-level: grammar-based method of creating vis \& Toolkits for Creating Immersive Analytics}
%   \item {gap: It is not intuitive for users to perceive and understand the construction process, and thus infer their analytical intention. }
  % To facilitate a better understanding and management in users visualization design process, some of the aforementioned systems provide history tool for users~\cite{heer2008graphical}. Just for recall ... low level click drags, high level filter, search...
  % Visualization tools and systems also help users to find the analytical logic and rationale of their visualization process.  We target facilitate understanding through embodied and intuitive creating process by embodied interaction in immersive env.

% interaction assists users to make use of information resources ~\cite{roth1997toward}, abstract nature, large quantities of information, diverse tasks
\textbf{Fluid Interaction for Immersive Analytics.}
Interaction plays a crucial role in visualization, serving as a key element in delivering an engaging user experience with visualized data~\cite{yi2007toward, elmqvist2011fluid,in2024evaluating}. This principle applies equally to immersive visualization~\cite{marriottCh2immersive, ens2021grand, mccormack2018multisensory}.
Following direct manipulation paradigms~\cite{shneiderman1983direct}, Elmqvist et al.~\cite{elmqvist2011fluid} introduced the notion of fluid interaction, emphasizing compelling and absorbing user experiences that maintain the flow of engaging in the tasks related to data visualizations~\cite{csikszentmihalyi1990flow}. 
Moreover, fluid interaction aims to reduce the gulfs of interaction—disparities between a user's intended actions and the system's provided affordances~\zq{~\cite{hutchins1985direct}.}

As immersive devices become increasingly prevalent, IA systems need to provide users with a fluid and directly manipulable experience~\cite{marriottCh2immersive, buschel2018interaction, lee2012beyond, roberts2014visualization}.
\zq{To facilitate data analysis with multiple views, Ens et al. proposed the \textit{Ethereal Planes} framework, which incorporates 2D information spaces into mixed reality environments~\cite{ens2014ethereal}. They introduced guidelines for interaction designers to create novel experiences with spatial interactions. 
Satriadi et al. investigated the design of multiview map visualizations in immersive environments, providing guidelines for IA and sensemaking through spatial interaction~\cite{satriadi2020maps}.} 
Bach et al.~\cite{bach2017hologram} investigated the effectiveness of direct manipulation, which is more aligned with interaction capabilities in immersive environments.
% Hayatpur et al.~\cite{hayatpur2020datahop} utilized the user's everyday spatial abilities to design a VR visualization system, implementing features such as riding an elevator to assist users in exploring and analyzing multidimensional datasets.
\zq{In addition, recent work by Lee et al. proposed a design space for the transformations between 2D and 3D views, considering natural and direct manipulation, such as the ``grab and pull'', to activate view transitions in mixed reality~\cite{lee2022design}. Another study explored the benefits of combining large interactive displays with personal head-mounted augmented reality for enhancing the exploration of superimposed views~\cite{reipschlager2020personal}.}
% The aforementioned works, \textit{Wizualization} and \textit{ImAxes}~\cite{batch2023wizualization, cordeil2017imaxes}, designed natural interactions to foster a fluid interaction experience.
However, previous works primarily focus on analytical tasks with immersive visualizations, neglecting the interactive experience related to creating composite visualizations. We argue that future IA systems with composite visualizations should consider a wider range of users and provide them with the experience of freely combining multiple visualization views in immersive environments.

%% file: sections/3_Concepts.tex
\section{Developing the Design Space: Key Considerations} \label{sec:concepts}
% \section{Developing the Design Space : Key Considerations} \label{sec:concepts}
The primary goal of this work is to involve users in the creation process of composite visualizations within immersive environments.
While previous research focused on the representations of composite visualizations in immersive environments, our focus lies in facilitating the creation process through natural and intuitive interactions. 
% This shift is based on the hypothesis that \textit{involving users in the creation process of composite visualizations by embodied interactions can yield a natural and fluid user experience in immersive environments.}
Our research is initially guided by the following questions:
\begin{itemize}
    \item \textbf{Q1.} What do we need to consider when combining multiple visualizations into a composite view?
    \item \textbf{Q2.} How can we design effective and fluid interactions to help users create composite visualizations in immersive environments?
\end{itemize}
To address \textbf{Q1}, we first introduce established categories of composite visualizations as our target for composition (\cref{spatialRelation}).
% combinable & detachable vis can be added here
Then, we identify the fundamental data relationships between visualization views that need to be considered when constructing a composite visualization. Subsequently, we associate the data relationships with the target composite visualizations to thoroughly explore their mappings (\cref{dataRelation}).
To answer \textbf{Q2}, we propose a \zq{schema} that incorporates data relationships and user interactions into the creation of composite visualizations.
% Then, we distill the design goals of the interactions essential for seamlessly constructing composite visualizations in immersive environments. 
We then develop a design space (\cref{sec:designspace}) and illustrate its usage in \cref{Guide4designspace}.
% In addition, we provide a series of interactive operations for building different types of composite visualizations, serving as a practical guide for using the design space .

% \begin{figure}[t]
% \centering
%   \includegraphics[width=0.99\columnwidth]{Figures/SpatialRelationship_ComVis.pdf}
%     \vspace{-2mm}
%   \caption{The five design patterns of composite visualization~\cite{javed2012exploring}, which is mainly based on the spatial relationship between visualization views.}
%     \vspace{-2mm}
%   \label{fig:CompositeVisViews}
% \end{figure}

\subsection{Spatial Relationship between Visualization Views} \label{spatialRelation}
Javed and Elmqvist introduced the notion of composite visualization and summarized five types of composite visualization~\cite{javed2012exploring}, including juxtaposed view, integrated view, superimposed view, overloaded view, and nested view. This categorization underscores the significance of spatial combinations of distinct visualization views. 
Accordingly, we transform our goal of creating composite visualizations into building these five spatial relationships. To illustrate these relationships, we use \textit{View A} and \textit{View B} as the primitive views that can be combined to form a composite view (\cref{fig:DataRelations}). 
% \zq{We further divide the spatial relationships into two categories, parallel and hierarchy relations based on whether there is a subordinate relationship  between \textit{View A} and \textit{View B}.}

\vspace{0.2em}\noindent{}\textbf{1. Juxtaposed views} involve presenting multiple views side by side with implicit linking in between. 
Representative examples include coordinated views and small multiples, extensively explored in visualization systems~\cite{roberts2007state} and immersive analytics environments~\cite{liu2020design}.

\noindent{}\textbf{2. Integrated views} share a similar visual composition with juxtaposed views but employ explicit linking, typically in the form of graphical lines. In Immersive Analytics, researchers have explored this by investigating the design space of drawing visual links~\cite{prouzeau2019visual,hubenschmid2021stream}.
% or creating the linked scatter plots~\cite{hubenschmid2021stream}.\\%Notable works is the \textit{VisLink}~\cite{collins2007vislink}.

\noindent{}\textbf{3. Superimposed views} overlay multiple visualization views atop one another to form a composite view. Early examples include Mapgets~\cite{voisard1995mapgets} and GeoSpace~\cite{lokuge1995geospace}, which overlay geographic visualizations with corresponding views to encode spatial relationships. Deng et al.~\cite{deng2022revisiting} comprehensively summarized examples of superimposed views. Yang et al. proposed immersive superimposed views based on map visualization, such as \textit{TiltMap}~\cite{yang2020tilt} and origin-destination flow maps~\cite{yang2018origin}.
% They are usually used to present the spatial relations between different visualizations. 

\noindent{}\textbf{4. Overloaded views} \zq{involve a client visualization overlaid on a host visualization without a one-to-one spatial linking between the two. Unlike superimposed views, overloaded views require modifications to the visual structures of the component visualizations rather than simply using visual layout operations to organize the views.}
% feature one host view and at least one client view, utilizing the same data encoding space as the host visualization. 
% Unlike other composite views, an overloaded view is formed by changing the visual structure rather than the layouts of views. 
Previous works on overloaded views include Scattering Points in Parallel Coordinates (SPPC)~\cite{yuan2009scattering} and the treemaps with overloaded graph links~\cite{Fekete_Wang_Dang_Plaisant_2003}. 
% In IA, there are relatively fewer examples of overloaded views~\cite{saffo2023unraveling}\zq{, such as 3De interactive lens~\cite{mota20183de}.}
In IA, examples of overloaded views are relatively rare~\cite{mota20183de, langner2021marvis}.
%Unlike superimposed or integrated views, there is no one-to-one mapping between the two visualization views.

\noindent{}\textbf{5. Nested views} also leverage the concept of host and client visualizations seen in overloaded views. However, in a nested view, the client visualization completely replaces the original visual component of the host view. Existing works mainly use graphs, matrices, or tables as the host views when designing nested views~\cite{deng2022revisiting}, such as \textit{NodeTrix}~\cite{henry2007nodetrix}, \textit{LSAView}~\cite{crossno2009lsaview} and the visual design in \textit{ProtoSteer}~\cite{ming2019protosteer}. Nested views in immersive environments have been relatively underexplored~\cite{saffo2023unraveling}.

We categorize these composite visualizations into \zq{parallel and hierarchical relationships} based on the presence of a subordinate relationship (i.e., host and client views) between views, as depicted in Fig.~\ref{fig:DataRelations}. 
% The parallel relationship encompasses juxtaposed, integrated, and superimposed views, where the two input visualizations are considered equivalent.
% The hierarchical relationship comprises overloaded and nested views, which involve visually embedding one or more visualizations into the encoding space of another visualization view.

\subsection{Data Relationships for Composite Visualizations} \label{dataRelation}
\begin{figure}[t]
\centering
  \includegraphics[width=0.99\columnwidth]{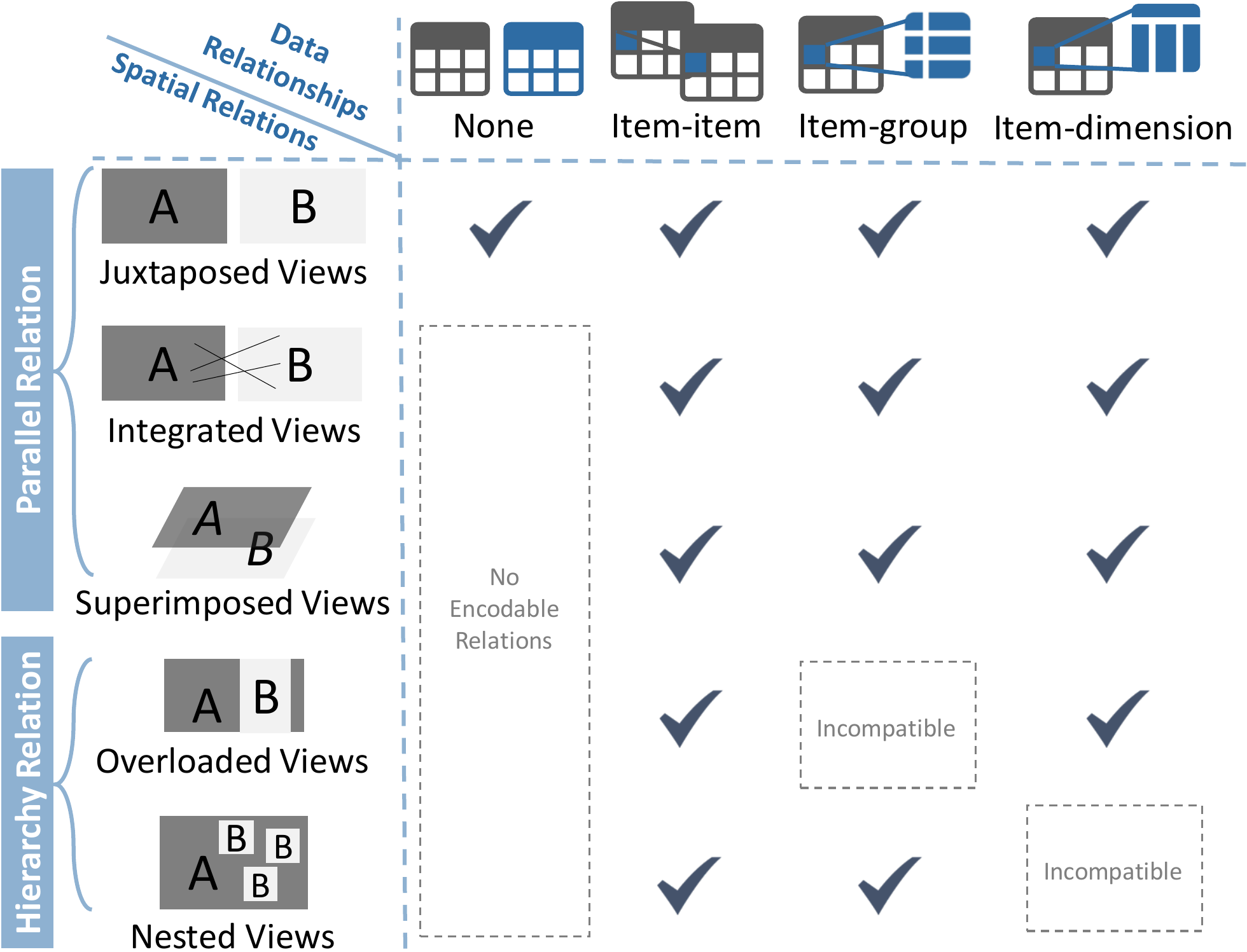}
    \vspace{-2mm}
  \caption{The constraints of underlying data relationships between different views on creating the five types of composite visualization.}
    \vspace{-2mm}
  \label{fig:DataRelations}
\end{figure}

The composition of multiple visualization views involves not only the spatial combinations but also the underlying data. 
% This can also serve as the foundation for more than two tables, as multiple tables can be merged into one single table. 
% We systematically enumerate all possible composite visualization types they can compose according to their data connections, as shown in Fig.~\ref{fig:DataRelations}.
Javed's work summarized four kinds of data relationships encoded by composite visualizations -- \textbf{\textit{None}}, \textbf{\textit{Item-item}}, \textbf{\textit{Item-group}}, and \textbf{\textit{Item-dimension}}~\cite{javed2012exploring}
We adopt and identify these relationships as the foundation and further summarize how the data relationship shapes the constraints on the types of composite visualizations that can be constructed between the two views (Fig.~\ref{fig:DataRelations}). 
We introduce the four data relationships as follows:
\begin{itemize}
    \item \textbf{None:} There is no overlap between the underlying data of the two data tables. 
    \item \textbf{Item-item:} There exists a one-to-one mapping of the data items between two tables.
    \item \textbf{Item-group:} The data relationship between the two tables is one-to-many, indicating that an item in one data table corresponds to multiple attributes of one item in the other data table (i.e., one row).
    \item \textbf{Item-dimension:} Simialr to \textbf{Item-group}, the data relationship between the two visualizations is one-to-many. The difference is that one item in one data table corresponds to multiple items under one certain attribute in the other table (i.e., one column).
\end{itemize}
%%%%%%%%%%%%%%%%%%%%%%%%%%%
%%%%%%%%%%%%%%%%%%%%%%%%%%%
%%%%%%%%%%%%%%%%%%%%%%%%%%%
% \begin{figure}[t]
% \centering
%   \includegraphics[width=0.99\columnwidth]{Figures/Formula.pdf}
%     \vspace{-2mm}
%   \caption{The high-level design framework that considers both data relationships and user interaction (user input) for constructing a composite visualization in immersive environments.}
%     \vspace{-2mm}
%   \label{fig:HighLevelFormulation}
% \end{figure}
%%%%%%%%%%%%%%%%%%%%%%%%%%%
% \subsection{Considering Both Data Relationships and Spatial Relationships Jointly} 
% \subsubsection{Connect Data Relationships to Spatial Combinations} 
%%%%%%%%%%%%%%%%%%%%%%%%%%%
\begin{figure*}[ht]
\centering
  \includegraphics[width=0.98\linewidth]{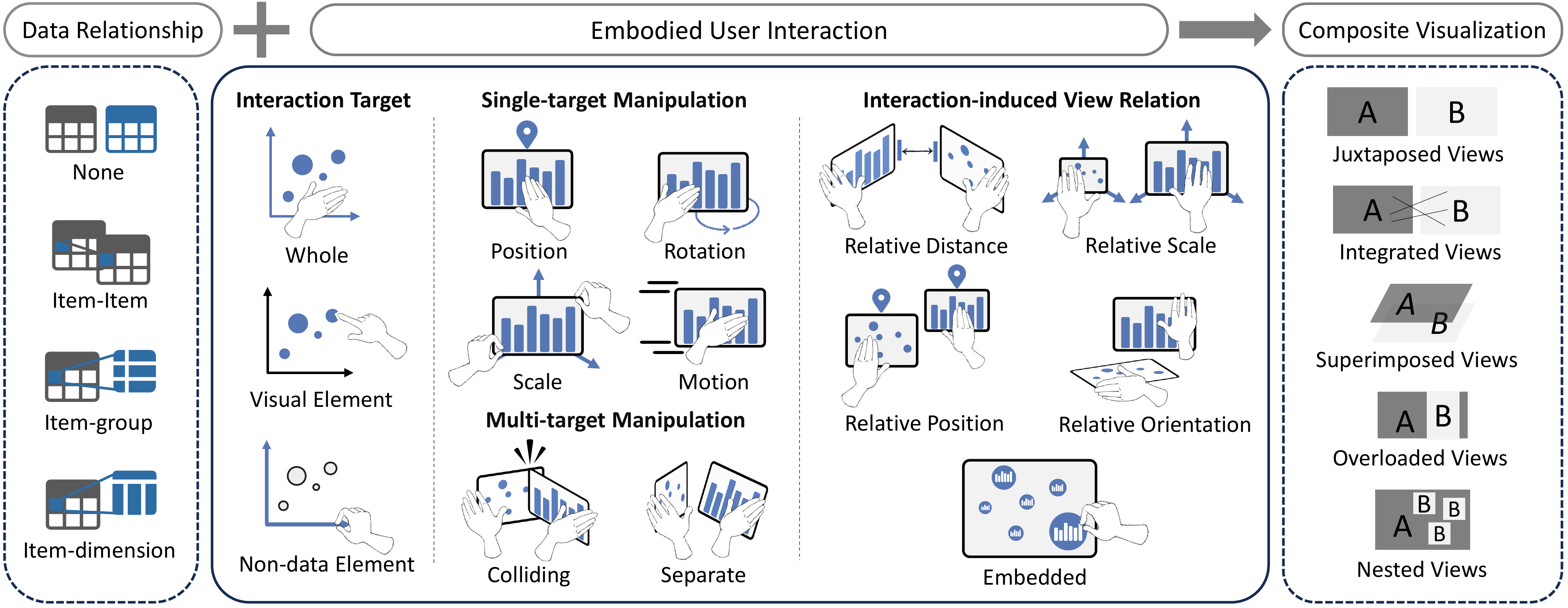}
    \vspace{-1mm}
  \caption{The design space of embodied compositing visualizations in immersive environments. It mainly introduces the embodied interactions for combining multiple visualization views to form a composite view, considering the constraints of the underlying data relationships.}
    % \vspace{-2mm}
  \label{fig:DesignSpace}
\end{figure*}
%%%%%%%%%%%%%%%%%%%%%%%%%%%

As shown in Fig.~\ref{fig:DataRelations}, we compile a table to exhaustively summarize the possibilities of creating all kinds of composite visualizations based on data relationships. We establish connections between the data relationships and the five types of composite visualizations.\\

\vspace{-0.8em}\noindent{}For \textbf{None} data relationship, no visual connections can be established between different views due to the absence of data association, allowing representation only through juxtaposed views.\\

\vspace{-0.8em}\noindent{}For the \textbf{Item-item}, juxtaposed, integrated, and superimposed views can be used to encode this relationship, as indicated in prior research~\cite{collins2007vislink, yang2018origin, liu2020design}. Overloaded views are also viable, as demonstrated in Yuan et al.~\cite{yuan2009scattering}, where elements in \textit{View B} (Fig.~\ref{fig:DataRelations}) can be mapped in a one-to-one manner to a portion of \textit{View A}'s data. While nested views typically involve one-to-many relationship~\cite{javed2012exploring, henry2007nodetrix} (as introduced in Sec.~\ref{spatialRelation}), they can also encode one-to-one data relationships in extreme cases (i.e., when \textit{View B} only encodes one data item).\\

\vspace{-0.8em}\noindent{}For the \textbf{Item-group}, juxtaposed or integrated views can be used to represent this relationship with visual links to present one-to-many connections. Superimposed views can also illustrate this relationship by stacking \textit{View A} above \textit{View B}, linking corresponding data regions~\cite{collins2007vislink}. Nested views can represent the item-group data relationship by replacing a component of \textit{View A} with \textit{View B}. However, overloaded views are not applicable here, as item-group involves one-to-many rows of a data table, whereas overloaded views correspond to \textit{View B}'s data being a subset of A's data rather than augmenting \textit{View A}'s data items (i.e., adding new rows in the data table of \textit{View A}).\\

\vspace{-0.8em}\noindent{}For the \textbf{Item-dimension}, similar to the item-group relationship, juxtaposed, integrated, and superimposed views are applicable. Overloaded views are feasible based on ~\cite{yuan2009scattering}. However, nested views can not represent this data relationship, as per their definition, where \textit{View B} replaces one or multiple visual elements in \textit{View A} without introducing new data attributes (i.e., adding new columns in the data table of \textit{View A}).
% View B visually replaces one or more data elements in View A, which encodes data items, rather than simultaneously replacing one or more attributes in View A's data.

In addition to data relations, it is also important to recognize that the design of composite visualization is not inherently unique from the perspective of visual encoding~\cite{yang2014understand}; rather, it depends on factors such as user needs, preferences, applicable tasks, and scenarios. 
Furthermore, in immersive environments, the rules of design and combination of data visual representations have not been fully explored~\cite{ens2021grand}. 
Therefore, we mainly consider objective data constraints rather than using visual design to constrain potential combinations between views.
% These constraints on data relationships can determine alternatives for creating composite visualizations. 
However, from the perspective of empowering users to actively participate in creating composite visualizations, we also need to match users' intentions to author different types of composite visualization from the perspective of interactions.
% none (no visual relation between different views), item-item (one-to-one visual relationship between items in different views), item-group (one-to-many relationship between items and groups), item-dimension (one view serving as the scale for another view). 
 % linked by showing different data aspects or for easy comparison of data or differences.
%not only add data relationships between overlap, but also item-all, entities description.

%% file: sections/4_DesignSpace.tex
\section{Design Space of Embodied Compositing Visualization} \label{sec:designspace}
% \section{Embodied Compositing Visualizations : Design Space} \label{sec:designspace}
\subsection{Immersive Visualization Compositing \zq{Schema}} \label{formula}
We construct a \zq{heuristic schema} for constructing composite visualizations with embodied interactions. 
As shown at the top of Fig.~\ref{fig:DesignSpace}, we consider the data relationships and user interaction to determine the resulting composite visualizations. 
% This uses the data relationships as the fundamental constraints and empowers users to create different compositions of visualizations through embodied interactions.
% Based on the constraints of data relationships on the left, users achieve the construction of different types of composite visualizations (on the far right) by combining intermediate interactive operations.
%------alternative to above------
This \zq{schema} prioritizes user-centric composition, allowing users to create diverse composite visualizations using intuitive interactions. With the data connections between views as objective constraints, users are not required to possess prior knowledge of the underlying data relationships; instead, the design space functions to infer the desired composite visualization based on user interactions.

\subsection{Interaction Design Rationale} 
We propose the interactions based on the following rationales:\\

\vspace{-0.8em}\noindent{}\textbf{DR1}: Design intuitive and \zq{easy-to-learn interactions for view composition.} \zq{To achieve this,} we follow the paradigm of direct manipulation~\cite{shneiderman1983direct}, which leverages familiar physical interactions (e.g., grasping and assembling) to ensure intuitive interactions.
\zq{Given the abstract nature of visualizations compared to physical entities and the complexity of the elements in one view, it is essential to divide a visualization into distinct interactive objects that correspond to different user intents.} 
We aim to \zq{reduce the learning curve for interactions and allow users to transfer their knowledge of spatial manipulations to an immersive environment with visualizations~\cite{marriottCh2immersive}.}\\
% Our aim is to ensure that the interactions are easy to learn while remaining consistent with their intuitive semantics for combining different views in immersive environments.\\
% Rapid, incremental, and reversible operations whose impact on the object of interest is immediately visible;
% Prompt feedback: users should be immediately informed of progress towards their goals;– Sense of control: ensure users feel in control over the activity so that they can truly affect the outcome; and

\vspace{-0.8em}\noindent{}\textbf{DR2}: Eliminate ambiguity in interactions to create \zq{distinct types of composite views.}
% Eliminate ambiguity between different interactions, allowing different operations to create the distinct types of composite visualizations.
To accurately convey user intentions and prevent errors in visualization composition, it is critical to design interactions that \zq{ensure clarity} and eliminate ambiguity.
\zq{For instance, when a user grabs a visualization view and places it in a specific position to indicate composition, these operations should result in a clearly defined type of composite view. Interaction designs need to be differentiated in terms of specific operations and interaction targets to ensure that different operations produce distinct outcomes, minimizing user confusion and enhancing the usability of interactions.}\\
% Mitigating ambiguity of interaction design is crucial to accurately convey user intentions when composing visualizations~\cite{norman2013design}. 
% The interactions should ensure that different sets of operations yield different outcomes and minimize confusion among users.\\ 
% We organze a series of interactive operations with each type of composite visualization interaction types 

\vspace{-0.8em}\noindent{}\textbf{DR3}: Provide \zq{smooth and} instant visual feedback for user interactions.
\zq{For view compositions, we assume that} users engage in a series of \zq{spatial manipulations with visualizations,} such as grabbing and assembling. 
To ensure a fluid experience~\cite{elmqvist2011fluid}, it is crucial for users to receive immediate visual feedback for each operation. 
\zq{For example, when a user selects an element in a view, we expect the system to instantly highlight the new configuration or selected effect. This feedback could confirm the user's action and help in understanding the impact of their interactions in real-time.}
% When creating composite visualizations, users engage in a series of embodied interactions such as grabbing, moving, and assembling. It is essential for users to receive immediate visual feedback from the manipulated visualization for each interactive operation to achieve fluid experience.

\subsection{Design Space Overview}  \label{designSpaceOverview}
We divide the embodied user interactions into three main components (as shown in \cref{fig:DesignSpace}): \textbf{(1) interaction target}, \textbf{(2) target manipulation}, and \textbf{(3) interaction-induced view relation} between two views.
Regarding the interaction target, we classify it into three distinct types, reflecting the typical components of a visualization view: 
(1) utilizing the \textbf{\textit{Entire view}} as the interaction target; 
 % \targetView{}
(2) engaging with a specific segment of \textbf{\textit{Visual Element}} within the visualization as the interaction target; and
% \targetVisual{}
(3) targeting \textbf{\textit{Non-data elements}} present in the visualization view, such as axes.
% \targetNon{}
For these targets, we outline a range of \zq{3D manipulations designed for operating with} either single or multiple targets. In Fig.~\ref{fig:DesignSpace}, we involve a single target view, users can modify its characteristics through four fundamental operations: changing its \textbf{\textit{Position}}, \textbf{\textit{Rotation}}, \textbf{\textit{Scale}}, and \textbf{\textit{Motion}}. 
When interacting with multiple target views, users can manipulate two of them simultaneously using both hands, either to \textbf{\textit{Collide}} them together or \textbf{\textit{Separate}} them from overlapping each other. 
\zq{All these manipulations are selected as fundamental and intuitive 3D manipulations to meet \textbf{DR1}. 
According to specific tasks or user requirements, designers can freely combine these basic 3D manipulations to design interactions for different targets to achieve \textbf{DR2}.
When choosing these manipulations, we did not limit input devices or modalities for generalizability. Considering the widely used input methods, such as controllers and hand gestures, these 3D manipulations can be activated by a single button on a controller to grab or release an object, or they can be driven by bare-hand interactions for grabbing and releasing an object.} 
% Designers and developers can flexibly combine these 3D  from our design space with specific requirements to achieve different interaction designs for creating composite visualizations.}
Following these manipulations, we analyze users' intentions by evaluating the interaction-induced states or relations between two or multiple views. The categories of interaction-induced view relations we consider include \textbf{\textit{Relative Distance}}, \textbf{\textit{Relative Scale}}, \textbf{\textit{Relative Position}}, \textbf{\textit{Relative Orientation}}, and whether the views are in an \textbf{\textit{Embedded}} configuration.
\zq{We offer smooth transitions and immediate visual feedback (\textbf{DR3}) when users interactively manipulate the views or elements to indicate the states or relations.}
\begin{figure*}[h]
\centering
  \includegraphics[width=0.98\linewidth]{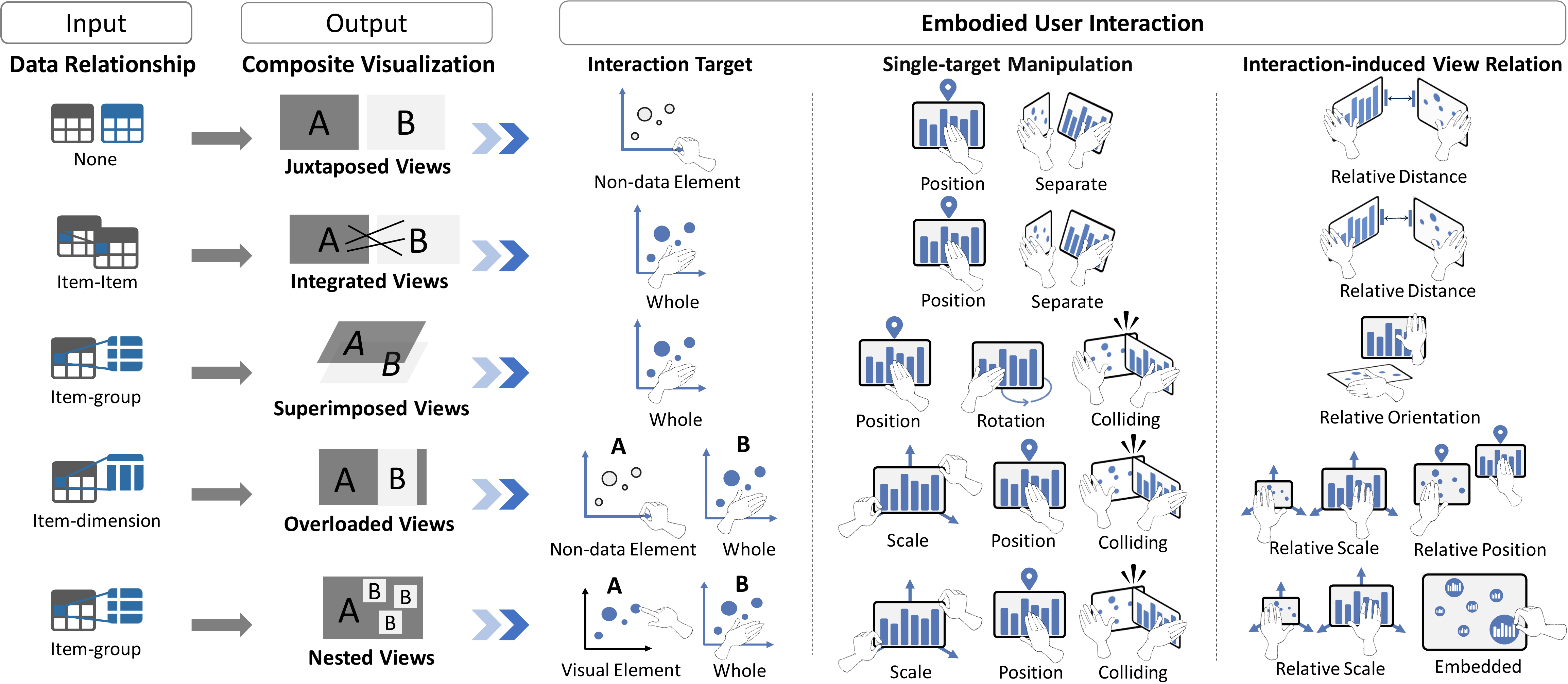}
    \vspace{-1mm}
  \caption{Illustration of using the design space for creating five \zq{examples} of composite views. \zq{We first determine data relationships as the input and the composite visualization to be created. Then, we design interactions by combining these 3D manipulations and assign them to different targets.}}
    \vspace{-1mm}
  \label{fig:DetailedManual}
\end{figure*}
%%%%%%%%%%%%%%%%%%%%%%%%%%%
% Based on the above design rationales, we proposed the design space of embodied interaction in the middle of .
% It contains three parts, interaction target, target manipualtion and .

\subsection{Using the Design Space} \label{Guide4designspace}
With the design space, we provide \zq{several examples to illustrate the design process of interactions} for creating five types of composite visualization.
% we establish a systematic linkage between data relationships, embodied interactions, and the specific types of composite visualizations to be created. 
As shown in Figure~\ref{fig:DetailedManual}, \zq{designers or developers can first determine the underlying data relations as the input. Then, they need to select the type of composite visualization to be created as the output. After determining the objective input and output type, they can select the fundamental 3D manipulations with interaction targets in the design space. They have the flexibility to adopt and combine various 3D manipulations with different interaction targets to propose intuitive and reasonable interaction designs.}
% we illustrate the process for crafting five types of composite visualizations based on a set of interactions. 
It is worth noting that the examples shown in Fig.~\ref{fig:DetailedManual} provide representative cases but may not be the only method suggested. 
We introduce the process of leveraging the design space to combine manipulations and propose specific interactions in the following examples.
% Designers or developers can adopt and combine various interaction technologies in the design space to tailor their systems for crafting composite visualizations according to the tasks and requirements.

\smallskip{}\noindent{}\textbf{Juxtaposed Views:} When the data relationship between visualization views is \textbf{\textit{None}}, users can create juxtaposed views by manipulating the \textbf{\textit{Position}} of \textbf{\textit{Non-data element}} in individual visualization views. For example, we could enable users to drag the x or y-axis to create new juxtaposed views around the original visualization view. We can also let them create juxtaposed views by putting the views side by side in a \textbf{\textit{Separate}} manner.
% For other types of data relationships, including \textbf{\textit{Item-item}}, \textbf{\textit{Item-group}}, or \textbf{\textit{Item-dimension}}, 
% a simple way for creating juxtaposed views is to put the views side by side in a \textbf{\textit{Separate}} manner.

\smallskip{}\noindent{}\textbf{Integrated Views:} To create integrated views, users can arrange multiple views together by manipulating the \textbf{\textit{Relative distance}} of them. When the distance between views is less than a certain threshold, and they do not collide or overlap (\textbf{\textit{Separate}}), users can create integrated views with explicit links. This method is similar to constructing the juxtaposed views, with the only difference being the inclusion of explicitly encoded data relationships. 
We recommend prioritizing the integrated view when an obvious data relationship exists between the views. 
% Alternatively, additional interactive operations could be introduced to enable users to filter or remove explicit visual links.

\smallskip{}\noindent{}\textbf{Superimposed Views:} Users can adjust the \textbf{\textit{Position}} and \textbf{\textit{Rotation}} angles of multiple views, \textbf{\textit{Colliding}} them together. For example, the user can create superimposed views by putting one view on top of the other at a specific angle and then blending them through collision.
This process involves analyzing the \textbf{\textit{Relative orientation}} of the two views when \textbf{\textit{Colliding}} them, which is the key factor for determining the construction of the corresponding superimposed view.

\smallskip{}\noindent{}\textbf{Overloaded Views:} For overloaded views, users can manipulate the host view (i.e., \textit{View A}) and the client view (i.e., \textit{View B}) separately. 
First, they can adjust the \textbf{\textit{Relative scale}} of them to create a noticeable difference in scale, indicating their intention for a host-client composition. Then, they can manipulate the \textbf{\textit{Non-data element}} of \textit{View A} to modify its visual structure to make space for \textit{View B}. For instance, users can change the visual structure of the parallel coordinates in SPPC~\cite{yuan2009scattering} by manipulating the axes. Then, they change \textbf{\textit{Position}} of \textit{View B} and \textbf{\textit{Colliding}} the two views convey their intention for composition.

\smallskip{}\noindent{}\textbf{Nested Views:} To construct nested views, users can adjust the \textbf{\textit{Relative scale}} of different views to ensure that the host view (i.e., \textit{View A}) scale is significantly larger than the client view (i.e., \textit{View B}). 
Then, the user can \textbf{\textit{Collide}} \textit{View B} with a visual component of \textit{View A} to express the intent of replacing the visual component by \textit{View B}. 
Once recognizing this user intention and the \textbf{\textit{Item-group}} data relationship, the system can allow the specified components of \textit{View B} and \textit{View A} to implement \textbf{\textit{Embedded}} state to generate nested views.

%% file: sections/5_Dev.tex
%Provide Rationale for Each Design and Examples!
%%%%%%%%%%%%%%%%%%%%%%%%%%%
\begin{figure}[t]
\centering
  \includegraphics[width=0.99\columnwidth]{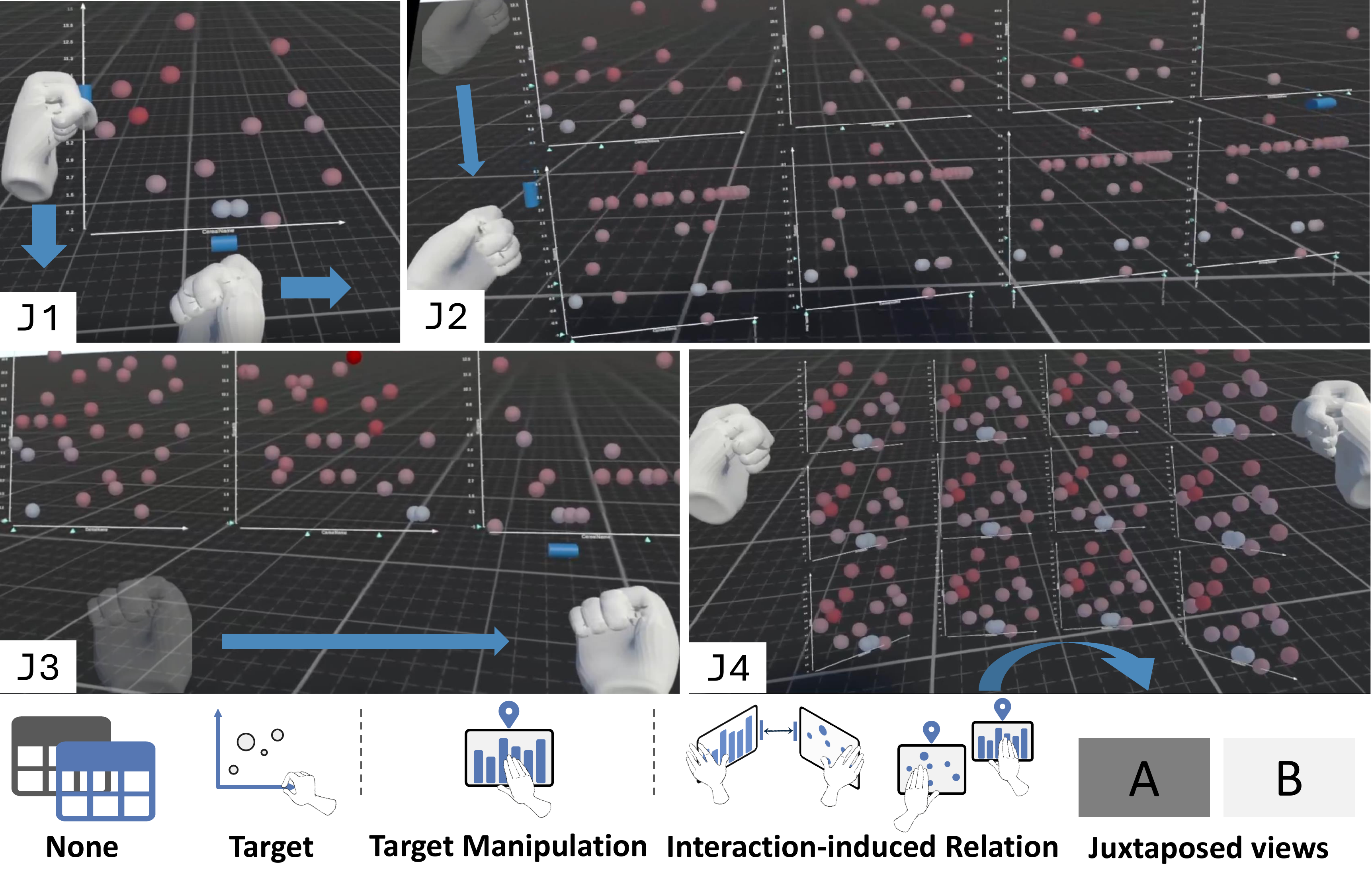}
    \vspace{-2mm}
  \caption{Illustration of creating juxtaposed views. 
  (J1) is the scatterplot that can be extended to small multiples by two embodied interactions: using bimanual interaction to extend the x and y axes at the same time in (J1), or using unimanual interaction to extend the y-axis (J2) vertically or the x-axis horizontally (J3). Users can also use both hands to grab and bend the small multiples to a desired curvature (J4).}
    \vspace{-2mm}
  \label{fig:JuxtaposedView}
\end{figure}
%%%%%%%%%%%%%%%%%%%%%%%%%%%
\section{Demonstration of the Design Space} \label{cases}
We demonstrate the design space through five proof-of-concept cases. They were implemented on Quest 3 using the Unity3D game engine and the Immersive Analytics Toolkit (IATK)~\cite{cordeil2019iatk}. 
We use them to demonstrate the interaction for creating each spatial composition type (\cref{spatialRelation}). \zq{Considering the three design rationales (\textbf{DR1-DR3}), we provide an intuitive interaction design for compositing views in each case by combining the intuitive manipulations in the design space (\textbf{DR1}). We carefully selected the interaction targets with reasonable 3D manipulations of each view to convey user intents for different compositions (\textbf{DR2}). We also provided smooth transitions with animations and visual transformations as immediate feedback of user interactions (\textbf{DR3}).}
% We illustrate each case by introducing the primitive visualizations we selected and then using a scenario to describe the interactions for the composition of the views. 
For cases where encoding relationships exist between different views (i.e., Case 2-4), we provide interactive techniques for decomposing views.
We recorded videos of these cases from both first- and third-person views as supplementary materials.
% in three general phases: (1) the primitive visualizations before creating the composite visualization, (2) the interactions to compose the primitive visualizations in a composite view, and (3) the created composite visualization and the corresponding interaction to break it down into the original views.

\subsection{Case 1: Juxtaposed Views}
We used small multiples as a representation of \textbf{\textit{None}} data relationships in \textbf{juxtaposed views}, as it is a commonly used visualization design, and the recent work has validated their advantages in immersive environments~\cite{liu2020design, liu2023datadancing}.
In this case, we allow users to \zq{embodied manipulate} the axes (\textbf{\textit{Non-data element}}) of the original visualization to create juxtaposed scatterplots. \zq{The usage of axes as interaction targets differentiates user intention of extending the view from manipulating the whole view (\textbf{DR2}).}
Users can perform both \textit{Expansion} and \textit{Partition} processing on the data of the original view through different interactive inputs with one hand or both hands \zq{directly (\textbf{DR1}).}
The interaction design was inspired by duplication tools that utilize copy-paste metaphors~\cite{liu2018data} to create multiple copies of the original view (e.g., \textit{Repeat Grid} in Adobe XD) and tools that segment views (e.g., \textit{Knife or Scissors} in Adobe Illustrator). 
\zq{We also provide animations as smooth transitions and real-time changes of views based on user interactions (\textbf{DR3}).}
% We illustrate the process of users creating small multiples by integrating a specific usage scenario.

\textbf{Usage Scenario:} %Data Extension and Partition \paragraph{Extension} \paragraph{Partition}
Kylie wants to understand and compare the sugar content of multiple cereal brands. She initially visualized the entire dataset in a scatterplot, as shown in \cref{fig:JuxtaposedView}-J1.
She was first interested in cereals with high sugar and low protein. She took down those brands and then wanted to explore more cereal brands in a wider range. Instead of creating a new chart, Kylie grasps the handlers on the x- and y-axis simultaneously (\cref{fig:JuxtaposedView}-J1). With a fluid motion, she expands the original data range and creates juxtaposed views that present a wider range of data. She can now explore various cereal brands distributed across different sugar and protein content intervals for comprehensive analysis. 
She then focused on cereal brands with high sugar and high protein content. However, that area was far from both axes, which was hard for Kylie to reference in a large scatterplot. Thus, she \zq{directly dragged} the handler on the y-axis to partition the data by sugar content, dividing the original data into multiple constituent views (Fig.~\ref{fig:JuxtaposedView}-J2). Then, as shown in Fig.~\ref{fig:JuxtaposedView}-J3, she does the same interaction on the x-axis to partition the data by protein content.
Furthermore, as shown in Fig.~\ref{fig:JuxtaposedView}-J4, she adjusted the curvature of small multiples by grasping the edges, demonstrating the unique advantage of embodied interaction in immersive environments
% This interaction embodies the unique advantage of embodied interaction in immersive environments, providing Kylie with the ability to customize the curvature of multi-view displays.

%%%%%%%%%%%%%%%%%%%%%%%%%%%
\begin{figure}[t]
\centering
  \includegraphics[width=0.99\columnwidth]{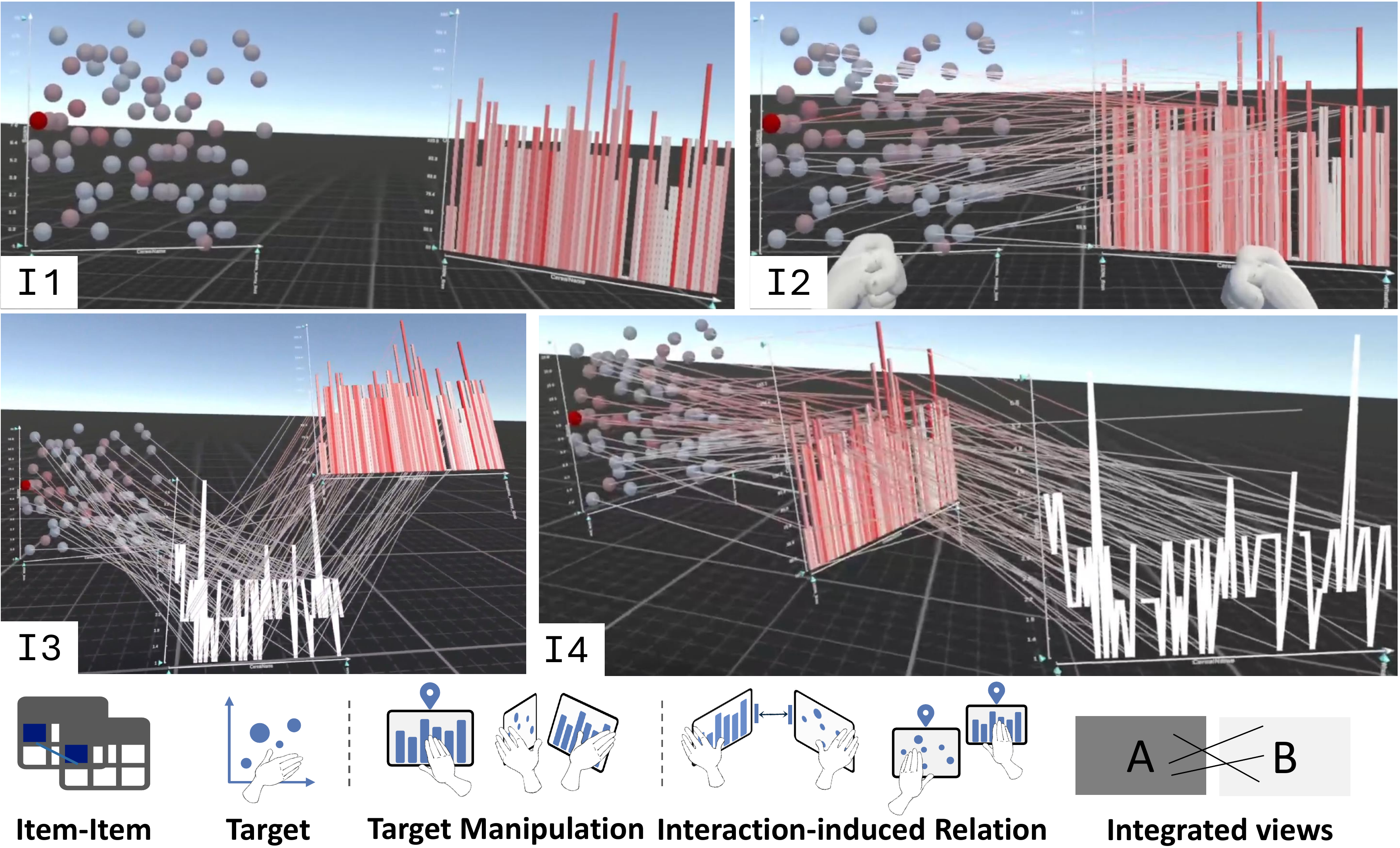}
    \vspace{-2mm}
  \caption{Illustration of creating integrated views. Users could grab and put two views in (I1) (i.e., a scatterplot and a bar chart) closely to compose an integrated view (I2). Users could also grab other views to create explicit links by manipulating their relative distance (I3) and adjusting their positions in immersive spaces (I4).}
    \vspace{-2mm}
  \label{fig:IntegratedView}
\end{figure}
%%%%%%%%%%%%%%%%%%%%%%%%%%%
\subsection{Case 2: Integrated Views} \label{integratedView}
% We utilized IATK to construct integrated views where multiple visualization views are connected by visual lines.
Graphical lines are commonly used in integrated views to express relationships between visualizations~\cite{javed2012exploring, collins2007vislink}. We implemented the case of integrated views that create visual links between different views based on embodied interactions of putting them closely.

\textbf{Usage Scenario}: As a product manager for a food company, Grace aims to understand the sugar, protein, and calorie content of different cereal brands. She presents the protein content using a line chart and the calorie content using a bar chart (Fig.~\ref{fig:IntegratedView}-I1).
Grace wants to identify cereal brands that are low in sugar and low in calories. Instead of identifying low-sugar brands and low-calorie brands and manually intersecting them, she picks up the views representing sugar content and calorie content and \zq{directly} brings them closer together (\textbf{DR1}). When the \textbf{\textit{Relative Distance}} between the views is less than a predefined threshold, it automatically generates graphical lines connecting the visual elements between them (\textbf{DR3}) (Fig.~\ref{fig:IntegratedView}-I2).
Furthermore, to additionally find cereal brands high in protein, Grace grabs the line chart and freely positions the three views close to each other \zq{without collision} (\textbf{DR2}) (Fig.~\ref{fig:IntegratedView}-I3 and -I4). 
She meticulously examines the data correlations from multiple perspectives and ultimately identifies the cereal brands she intends to choose. After that, Grace wants to examine the protein content individually in a line chart. She grabs the line chart and moved it away from the other views to easily decompose the integrated views.

% transition design of creating composite vis
\subsection{Case 3: Superimposed Views}
\begin{figure}[t]
\centering
  \includegraphics[width=0.99\columnwidth]{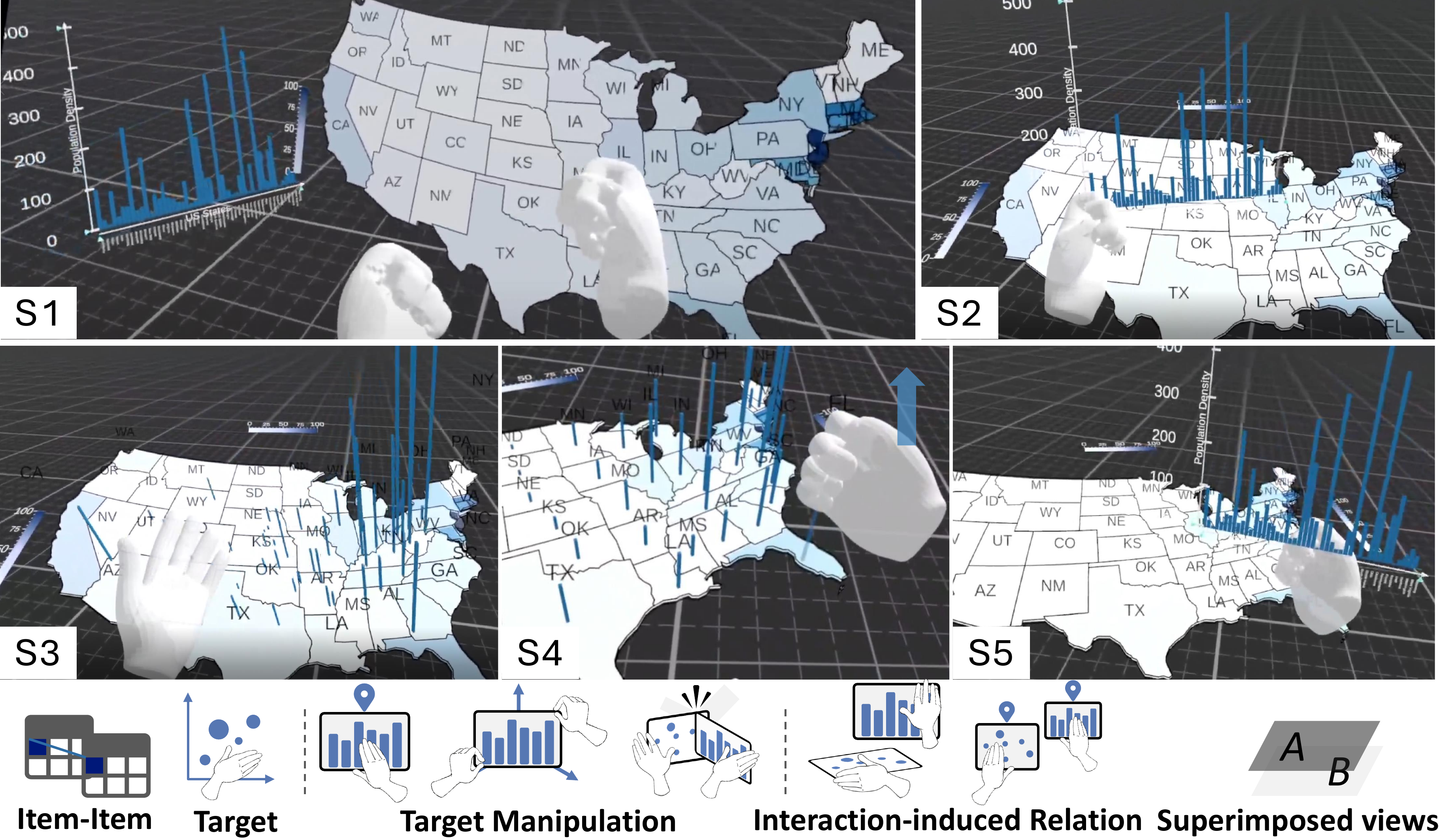}
    \vspace{-2mm}
  \caption{Illustration of creating superimposed views. Users combine a bar chart and a map (S1) by adjusting their relative positions and colliding them in relatively vertical orientations (S2). Then, the bars will spread out to the corresponding areas of the map (S3) to form a superimposed view. Users could decompose the visualization by grabbing a bar and pulling it up to separate the two views (S4-S5).
  }
    \vspace{-2mm}
  \label{fig:SuperimposedView}
\end{figure}

We drew inspiration from the \textit{Tiltmap}~\cite{yang2020tilt} and created a case of superimposed views based on a combination of maps and bar charts. 
% This case takes into consideration that in immersive environments, human perception is more sensitive to depth than color~\cite{munzner2014visualization}.
In Fig.~\ref{fig:SuperimposedView}-S1, we mapped the population density of each state in the United States using a map, with varying shades of color indicating density. Also, we had a bar chart showing the population density of each state. Other data properties could also be used. Below is a specific usage scenario to illustrate the interactions.

\textbf{Usage Scenario}: Ben aims to visually compare the population density of different states in the U.S. 
When examining the map, he struggled to distinguish the population density differences in the central regions due to similar colors among these states. Turning to the bar chart for data comparison, he faced the challenge of frequent switches between the map and the bar chart, as the bar chart lacked spatial context. 
To address this, Ben lifts the bar chart and positions it vertically above the map (\textbf{DR2}) (Fig.~\ref{fig:SuperimposedView}-S2). 
Such interaction generates a composite visualization by overlaying the bars onto the map, with an animated transition of spreading out the bars to the corresponding states (\textbf{DR3}) (Fig.~\ref{fig:SuperimposedView}-S3). 
After analyzing the population density data, Ben wanted to examine the bar chart separately to see the ranking of his home state in population density. He quickly lifts a bar by hand (\textbf{DR1}) (Fig.~\ref{fig:SuperimposedView}-S4), seamlessly returning the original map and bar chart (\textbf{DR3}).
% After analyzing the population density data, Ben desires to examine the state of his homeland outlined on the map. He quickly lifts any bar by hand (Fig.~\ref{fig:SuperimposedView}-S4), seamlessly returning the composite visualization to its original map and bar chart components.
%%%%%%%%%%%%%%%%%%%%%%%%%%%
%%%%%%%%%%%%%%%%%%%%%%%%%%%

\subsection{Case 4: Overloaded Views}
We present a representative case of overloaded views with Scattering Points in Parallel Coordinates (SPPC)~\cite{javed2012exploring}. We implemented a VR version of the SPPC based on the data processing and representation algorithms described in~\cite{yuan2009scattering}.

\begin{figure}[t]
\centering
  \includegraphics[width=0.99\columnwidth]{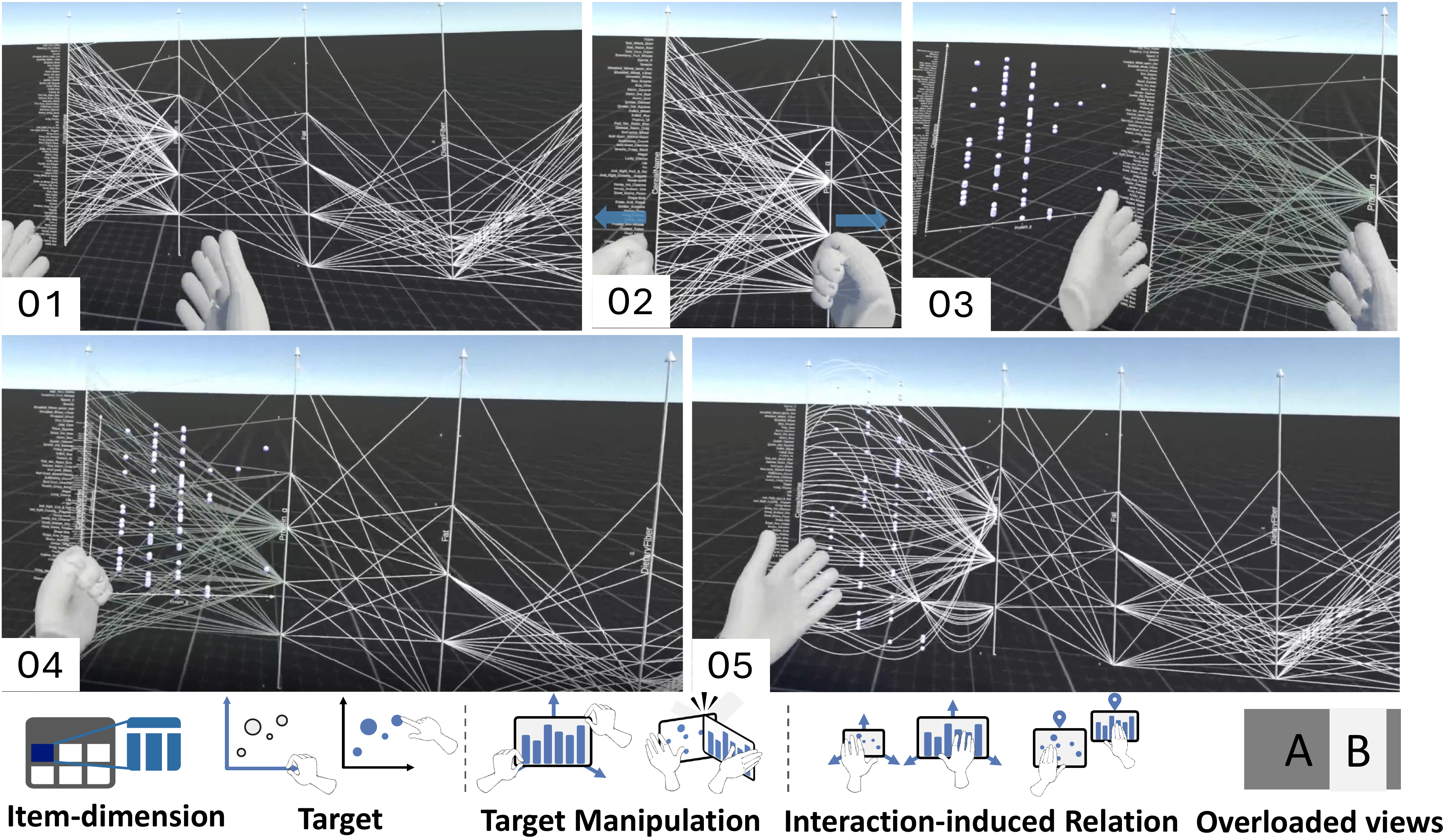}
    \vspace{-2mm}
  \caption{Illustration of creating overloaded views. Taking the Parallel Coordinated Scatterplot (PCP)~\cite{yuan2009scattering} as the host view (O1), users spread apart any two adjacent parallel axes with both hands (O2). The corresponding client view, a scatterplot, which represents the data between the two axes will appear adjacent to them (O3). Next, users place the scatterplot into the designated area of the PCP, thus creating overloaded views.}
    \vspace{0.2em}
  \label{fig:OverloadedView}
\end{figure}

\textbf{Usage Scenario}: Jessica, a nutritionist, is currently evaluating and analyzing the nutritional components of over 30 types of cereals. In front of her is a Parallel Coordinates Plot (PCP) visualization, as shown in Fig.~\ref{fig:OverloadedView}-O1), depicting the names of these cereals along with their sugar, protein, calorie, and dietary fiber content. She examines the parallel axes in the PCP and observes the connections between multiple nutritional attributes of cereals represented by graphical lines. 
However, she found that these lines were dense, hindering her ability to perceive data correlations effectively. 
If the data from two parallel axes in the PCP were plotted using scatterplots, it would visually demonstrate clear clusters and distribution patterns of the data to complement this issue~\cite{yuan2009scattering}. 
Therefore, Jessica would like to combine the subviews between adjacent axes in the PCP with scatter points so that she can effectively leverage the advantages of both visual representations.
To achieve this, she grabs each of the two axes with her hands and opens them apart \zq{to indicate her intention of selecting the specific area in a host view} (Fig.~\ref{fig:OverloadedView}-O2). The lines between the two axes become highlighted, and the corresponding scatterplot appears beside them (\textbf{DR3}) (Fig.~\ref{fig:OverloadedView}-O3). She then picks up the scatterplot \zq{(client view)} and places it within the highlighted area in the PCP to indicate the composition (\textbf{DR2}) (Fig.~\ref{fig:OverloadedView}-O4).
The immersive analytics (IA) system identifies the corresponding data relationships between the two views and then automatically generates a composite visualization consisting of overloaded views, as depicted in Fig.~\ref{fig:OverloadedView}-O5. 
In this way, Jessica can analyze the distribution of sugar content among different cereals. She also interacts with the other axes of the PCP in the same manner to examine the correlations and data distribution among different nutritional components. 
When she wants to view the PCP data lines or scatter points separately, she grabs the two axes again to bring them closer together (\textbf{DR1}), thus decomposing the overloaded views back into the original PCP.
%%%%%%%%%%%%%%%%%%%%%%%%%%%
%%%%%%%%%%%%%%%%%%%%%%%%%%%

\subsection{Case 5: Nested Views}
\begin{figure}[t]
\centering
  \includegraphics[width=0.99\columnwidth]{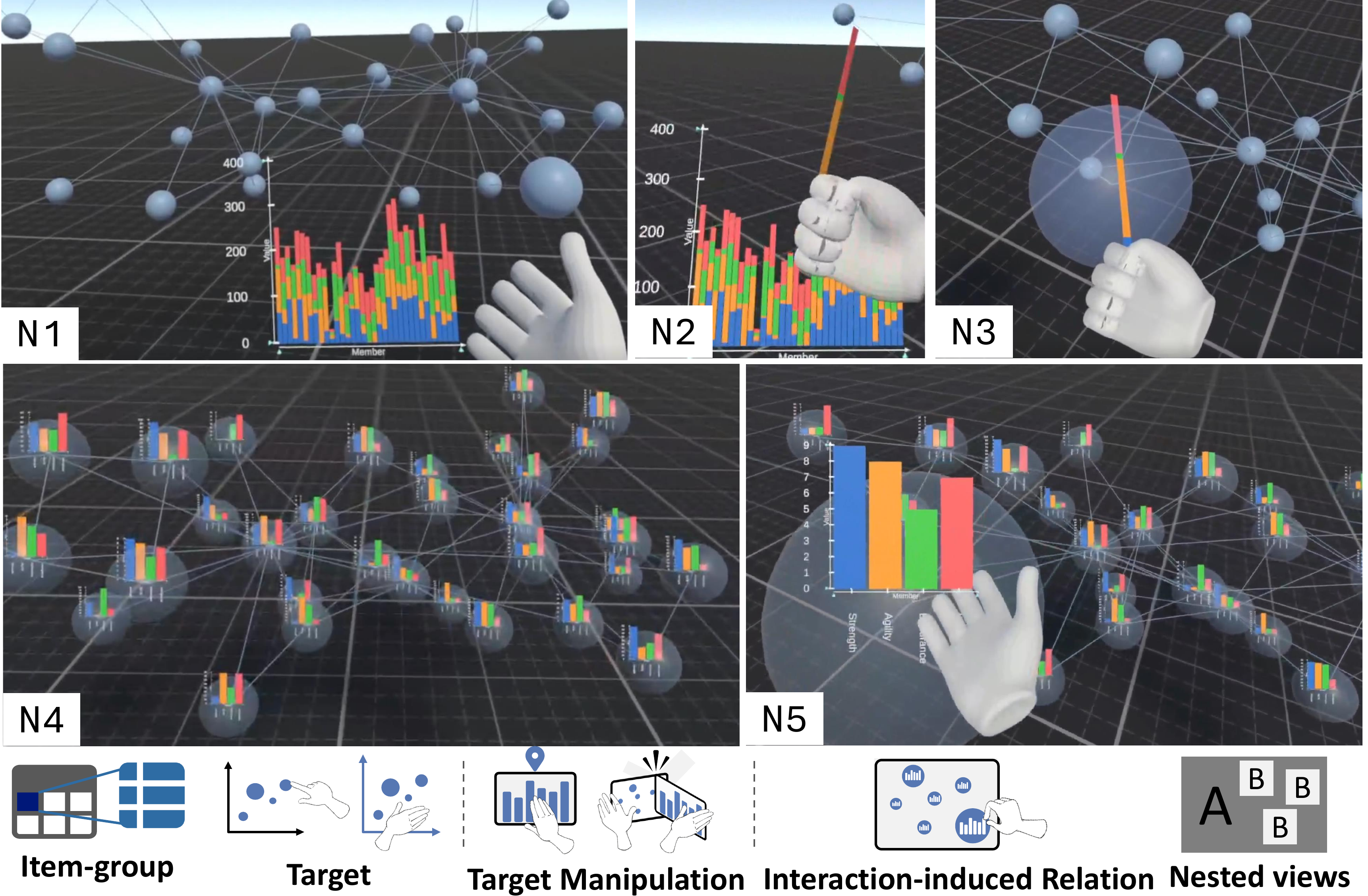}
    \vspace{-2mm}
  \caption{Illustration of creating nested views. We take the graph as a host view and a stacked bar chart as the client view (N1). After performing a pull-out gesture on a stacked bar chart to extract a single bar (N2), users can then place it into any node of the graph to nest it into the node (N3). The system will automatically match the other stacked bars with nodes and generate the nested views smoothly (N4). Users can reach their hands into any node to hover over the detailed information (N5).
  }
    \vspace{-2mm}
  \label{fig:NestedView}
\end{figure}

% We designed this case by drawing inspiration from 
This case is inspired by the representative nested views in composite visualization~\cite{javed2012exploring}, which combines a graph with a bar chart. We chose graphs as they exhibit inherent advantages when presented in immersive environments~\cite{kwon2016study,huang2023embodied}.
In Fig.~\ref{fig:NestedView}, we created a graph with the data of each node visualized by a bar chart.

\textbf{Usage Scenario}: John, as a Taekwondo enthusiast, aims to analyze combat data of different players in a Taekwondo fighting game. He utilizes graph data where each node represents one player, and the edges denote matches between them. Each player has four attribute values, including strength, agility, endurance, and intelligence, represented by stacked bar charts in Fig.~\ref{fig:NestedView}-N1. John wants to compare the agility values of two players in a match. 
However, it is challenging to compare them directly from the stacked bars because the different bars were not aligned, and he also needs to examine each player's competitors. Therefore, he plans to merge the graph and stacked bars into a composite view to facilitate the comparison of various attribute values of combatants in different matches.
% He decides to merge the stacked bars representing each player with their corresponding nodes. 
John begins by performing an extraction action on the stacked bar chart to get a stacked bar representation (Fig.~\ref{fig:NestedView}-N2). Then, he places it into one node of the graph visualization to indicate his intention to construct the nested view (\textbf{DR1}) (Fig.~\ref{fig:NestedView}-N3). 
Recognizing the collision and embedded states between the grabbed bar and the node, the IA system smoothly animates the corresponding stacked bars into the respective nodes (\textbf{DR3}) (Fig.~\ref{fig:NestedView}-N4). 
As shown in Fig.~\ref{fig:NestedView}-N5, in the created nested views, John touches each node to view and compare the specific attribute values of different players.
If he wants to view the original stacked bars to compare the overall strengths of multiple players, John could grab the bar chart from any node. The composition of a graph and stacked bars is then decomposed back into the original views.
%%%%%%%%%%%%%%%%%%%%%%%%%%%
%%%%%%%%%%%%%%%%%%%%%%%%%%%

%% file: sections/6_Study.tex
\section{User Study}
We conduct a user study to (1) \zq{assess the usability of the interaction design with the pre-designed cases and (2) gather insights on the user experience of the interactive creation of composite visualizations in immersive environments.}
% (3) Evaluate the practical application and utility of the design space by observing how participants use it to create composite visualizations.
% evaluate the usability of the interaction design based on the representative cases and (2) assess the user experience of creating composite visualization through embodied interactions in immersive environments.

\subsection{Experimental Setup and Participants}
Following Institutional Review Board (IRB) approval, our study investigated user interactions utilizing the Meta Quest 3 virtual reality headset, which features a resolution of 2064 x 2208 per eye and a refresh rate of 90 Hz. We leveraged the Air Link feature provided by Meta to offer a wireless experience, while still harnessing the computational power of a connected PC. \zq{During the study, we chose controllers as the input device to ensure stability. Participants only used the grip buttons of both controllers to indicate grabbing or releasing objects for 3D manipulation.}
The experimental setup was contained within a 3 x 2.5 meter space, affording participants the freedom to navigate within this designated area. At the commencement of each session, participants were guided to the center of this space to begin their experience.

We recruited 16 participants by sending recruitment advertisements via mailing lists and social media at a local university \zq{(ages ranging from 23 to 32, 7 females).} Participants included six individuals with expertise in visualization and five with expertise in extended reality (XR). Two of them have expertise in both XR and visualization. The other participants were graduate students from a variety of disciplines, including but not limited to computer science, finance, design, cybersecurity, and aerospace engineering.
% They all have the potential need to understand, analyze, and communicate data using composite visualization in their lives. 
% We recruited both experts and non-experts to gather their feedback on required composite visualization for their work.% we focused on popular productivity-related or day-to-day tasks that everyday consumers could find relevant and we avoided niche gaming applications

\subsection{Study Design and Procedure}
The study was structured into four phases, with two co-authors as experimenters throughout the study.

\noindent{}\textbf{1. Introduction:} We began by having participants read and sign a consent form. Then, we introduced them to the concept of composite visualization, covering background, definitions, and showcasing different types of composite visualizations (Fig.~\ref{fig:DataRelations}).
% We explained the design space (Fig.~\ref{fig:DesignSpace}) and the interactions, including interaction targeting, single-target manipulation, and multi-target manipulation.

\noindent{}\textbf{2. Training:} We let the participants enter a VR scene and briefly introduced the interactions in Fig.~\ref{fig:DesignSpace}. \zq{We presented the design space to guide them to familiarize themselves with all the basic interaction operations through embodied experiences with pre-created cubes (e.g., scaling a cube or colliding two cubes). This served as foundational training} for their subsequent experience with the pre-defined cases.
% and let them familiarize themselves  

\noindent{}\textbf{3. Experiencing interactions:} We invited them to experience the implemented cases (Sec.~\ref{cases}) in VR one by one. 
% We provided transitions between each case, allowing participants to freely switch to different cases.
At the commencement of each case, participants were presented with two or more primitive visualizations in the VR environment, poised for composition. Experimenters verbally introduced the dataset and the application scenario relevant to each case and guided the participants to construct the designated composite visualizations through embodied interactions. 
% This process required participants to utilize solely a ``grab'' operation, facilitated by a single button on the VR controller, thereby eliminating the need for memorization of any additional functionalities.
% They encountered two or more primitive visualizations as the initialization of each case and were verbally instructed to create composite visualizations using interactions in the design space. During this process, the experimenter introduced the dataset and usage scenario of each case and succinctly conveyed the required operations that the participants needed to perform. We encouraged them to think aloud and express their thoughts during the study. 
% We did not impose any time restrictions on the experiment, allowing users to freely explore each case and attempt the interaction operations repeatedly. 
The entire process lasted approximately 30 minutes within the VR environment, encompassing both the training and think-aloud, as well as the time allocated for participants to freely explore the cases.

\noindent{}\textbf{4. Questionnaire and Interview:} At the end of the study, participants filled out the questionnaires and participated in an interview.

\subsection{Data Collection}
In the third phase of the study, one experimenter monitored the participants' actions through the screen cast of the VR environment, while the other experimenter documented observations related to any challenges participants encountered, instances of confusion, or notable instances of creativity exhibited during the interaction. 
The entire VR study, including participants' verbal feedback and the VR scenes, were recorded using an external camera, as well as the VR view cast.

In the fourth step, we collected subjective ratings on the usability and user experience of performing the embodied interactions for composing views.  Our questionnaire for evaluating usability and user experience was inspired by established frameworks in the literature~\cite{nielsen1994usability, o2010development}. 
Usability assessment focused on dimensions such as learnability, efficiency, memorability, sense of control, and overall satisfaction with the interaction. To evaluate the interaction experience, we collected participants' ratings of the interactive experience by the naturalness, consistency with real-life operations, engagement, and enjoyment~\cite{hung2017assessing}.

\subsection{Results}
\textbf{Usability.} All participants successfully constructed composite visualizations in the provided cases. Generally, they perceived the interactions as easy to learn ($MEAN = 6.3, SD = 0.8$) and remember ($MEAN = 6.2, SD = 0.8$). 
\zq{Nine} participants remarked that the interactions \textit{``felt intuitive (P3)''} and \textit{``required minimal effort to understand (P15).''}
% More than half of the participants (9/16) mentioned that \textit{``these interactions are very intuitive (P3)''} and \textit{``no need to spend time memorizing or learning, I can figure them out in a short time (P15).''}
% In addition, most participants (11/16) perceived these interactions as effective ($MEAN = 6.4, SD = 0.6$) and satisfied ($MEAN = 6.0, SD = 0.6$) because they reported that the interactive operations align with their cognition or \textit{``they match my thinking (P9).''} 
Moreover, they perceived the interactions as effective ($MEAN = 6.4, SD = 0.6$) and satisfactory ($MEAN = 6.0, SD = 0.6$), as \zq{eleven} of them felt the interactions were aligned with their mental model or \textit{``matched my thinking (P9).''} 
Five participants stated that the interactions corresponded well with the semantic concept of composite visualization. 
For example, one participant mentioned that \textit{``It is very straightforward to create the [superimposed] view by grabbing and placing one view on top of another (P6).''} 
Overall, participants rated the interactions as controllable ($MEAN = 5.3, SD = 1.0$), noting that \textit{``the dynamic interactions are smooth (P7).''} However, three participants were concerned about unintended merging views.
% However, three of them expressed concern about the potential for erroneous merging of views.

\textbf{User Experience.}
For the embodied interaction experience, all participants perceived the interaction to be natural ($MEAN = 5.5, SD = 1.0$), engaging ($MEAN = 6.4, SD = 0.7$), and enjoyable ($MEAN = 6.7, SD = 0.6$). 
Based on the feedback from interviews, we found that the interactions potentially promoted user experience in three aspects: 
(1) providing visualizations with a sense of \textbf{physical affordance}, making the interaction with data visualizations more akin to manipulating physical objects;
(2) facilitating the \textbf{comprehension of relationships} between different data visualization views; and
(3) offering \textbf{flexibility} in analyzing data views through both separated and composed views.
% Furthermore, nearly half of the participants emphasized the importance of \textbf{integrating the interactive creation process with analytical tasks} to ensure a engaging and smooth data analysis experience.

\textbf{Physical Affordance.} More than half of the participants noted that the interactive experience resembled physical interactions in daily life, such as pulling, dragging, and lifting. They highlighted that this interaction \textit{``transforms visualizations from abstract data into physical entities (P1).''} 
Three participants even drew parallels between the visualizations and physical objects. For instance, they likened the map to \textit{``a piece of paper (P7)''}, or described the nodes in the graph as \textit{``floating balloons (P5).''} 
In addition, five participants emphasized the engagement offered by this interactive experience, particularly appreciating the ability to simultaneously manipulate visualizations with both hands, stating that it \textit{``gives me the feeling like I am actually playing with the data (P4).''}
From the perspective of a financial data analysis expert, one participant valued this interactive experience highly, likening the VR experience to a game-like scenario and finding it enjoyable. 
The participant even expressed willingness to pay for such an experience in the future, comparing it to \textit{``purchasing LEGO sets (P9).''}
However, three participants expressed concerns about the practical application of this interactive experience with composite visualizations.
They suggested that \textit{``additional complex functionalities might be necessary (P16)''} to aid in the process from constructing to analyzing data visualizations. They also raised concerns that overly complex interaction designs could potentially lead to confusion and increase the learning curve.

\textbf{Comprehension of View Relations.} \zq{Twelve of the participants} indicated that combining composite visualizations through embodied interactions intuitively facilitated their understanding of the relationships between different views. According to one participant, \textit{``with this hands-on approach, I grasp how these views are connected (P7).''}
% , thus comprehending the relations encoded by composite visualization. 
In addition, two visualization experts (P5, P13) mentioned that reading composite visualizations is often more challenging for non-visualization experts, and this interactive approach may \textit{``enhance visualization literacy among everyday users (P13).''}
However, three participants pointed out that they cannot rationalize their intent for performing these interactions without a prior in-depth understanding of the semantic relationships between the views. They emphasized the need to \textit{``incorporate these interactions into tasks related to relationship analysis (P16).''} 
Similarly, five participants expressed a desire for specific interactions to filter or highlight specific data in composite visualizations, although our study primarily focused on the creation process. 
Furthermore, they emphasized the importance of incorporating precise analytical intent as motivation for constructing composite visualizations. For instance, when creating the integrated views in \cref{integratedView}), four participants expressed a desire to \textit{``select specific data and create visual connections only between those items (P5).''}

\textbf{Flexibility of Analyzing and Communicating Data.} \zq{Fourteen of the participants} appreciated the capability and flexibility provided by composing and decomposing views for two reasons.
First, they expressed appreciation for the freedom to \textit{``switch between the separate and composite views (P6)''}, which offers potential efficiency and flexibility for understanding data.
In addition, all visualization experts pointed out that this flexibility may facilitate Exploratory Data Analysis (EDA) because they reported that the ability to freely combine and split views allows analysts to \textit{``validate hypotheses by composing a new view or splitting one into multiple parts (P15).''}
Second, fourteen of them indicated that the freedom of combining views is well-suited for presentations or communication scenarios with visualizations. 
The reason is that they thought live demonstrations of such combined visualization views can \textit{``present ideas or persuade others in a more compelling and engaging way (P14).''}
However, two participants mentioned that this flexibility also puts forward higher requirements for the consistency of interaction design. For example, users may need \textit{``exactly corresponding operations (P7)''} to combine and split visualizations.
% they felt \textit{``treating the visualization as controllable entities''}. 

% \textbf{The Need of Integrating Creation and Analysis Workflows.} During the study, 
% Furthermore, they expressed the desire to articulate their precise analytical intentions during the visualization creation phase. For instance, in the creation of integrated views (Sec.~\ref{integratedView}), seven participants would like to \textit{``select the specific data I am interested in and then create the visual lines only between these data subsets (P5)''}.

%% file: sections/7_Discussion.tex
\section{Discussion}
Our work employs two metaphors to conceptualize composite visualization.
First, we envision visualization views as composable and detachable modules, thereby turning creating composite visualizations into a process similar to physical assembly.
Second, we combine a series of 3D manipulations for crafting interactions, which allow users to assemble and disassemble different types of composite views naturally.
Based on this, our work can provide fresh perspectives for shaping interactive experiences with composite visualizations in immersive spaces.
We reflect on our case implementations and study findings by discussing: 1) usability, 2) physical affordance, 3) integration of visualization creation and analysis, and 4) facilitating flexibility. Then, we discuss future work about generating composite visualizations and the potential usage scenarios.
% design considerations and key questions regarding active engagements with composite visualizations, as well as relevant usage scenarios in immersive environments.
% Through the implementation of the exemplary cases and the user study, we identified design considerations and key questions for inveloving active experience with composite visualization, as well as the potential usage scenarios that may be beneficial for promoting user experience with composite visualizations in immersive environments.

\zq{
\textbf{Usability of Interaction Design.} Our study demonstrates the benefits of intuitive interactions for creating composite views. This aligns with previous research that emphasizes the importance of intuitive interaction~\cite{lee2022design, bach2017hologram}. 
Our interaction design mainly relies on various combinations of basic operations to clearly express users' intentions to compose views. This method minimizes the complexity of interactions, making it easier for users to learn. However, this combinatory approach may lead to overlapping operations between different interaction designs, potentially resulting in the unintended composition of views or the accidental activation of unwanted operations. Therefore, it is necessary to highlight the differences between interactions to accommodate more complex user commands for manipulating composite views.
}

\textbf{Designing Interactions Using Familiar Physical Metaphors.}
\zq{Our study validates the intuitiveness and effectiveness} of using physical metaphors to design immersive visualization interactions, resonating with previous research~\cite{cordeil2017imaxes}.
However, our work represents only an initial exploration of these metaphors for authoring composite views. 
% \zq{When considering the subsequent data analysis or more complex tasks,} the intuitive and natural operations need to be combined for more precise commands. 
\zq{To extend the current space for data analysis tasks,} Immersive Analytics (IA) developers or designers may leverage other real-life metaphors, such as incorporating physical tools as interaction controls. 
% There are still several factors that needs to be considered \zq{when designing interactions for analyzing immersive visualizations.}
% First, the intuitive and natrual operations need to map to precise interaction commands for data analysis. 
For example, if a user needs to delete an element, they could grab and throw it~\cite{in2023table} or put it into a virtual trash bin to convey the delete operation.
However, when the number of composite views increases, participants may struggle to keep track of each view. Therefore, it is necessary to incorporate other types of feedback beyond the visual channel, such as haptic feedback~\cite{kraus2020assessing}. This could provide users with spatial perception or guidance \zq{to accommodate complex user inputs for data analysis}.
% combine natural metaphor-based interaction with the WIMP user interfaces for manipualting composite visualization.
% In addition, we could also  
% In immersive spaces, people tend to pursue natural interfaces and have elevated requirements, such as using minimal buttons or menus while demanding precise input and clear operation instructions~\cite{lee2021post}. 

\textbf{Integrating Visualization Creation and Analysis \zq{Workflow}.}
Previous research underscored the advantages of employing natural, intuitive interactions for analyzing composite visualizations in IA~\cite{yang2020tilt, yang2018origin}. Our \zq{findings indicate that user engagement in the creation of composite visualizations may facilitate a seamless transition from visualization creation to subsequent data analysis. 
However, the intuitive and easy-to-use interactions, due to their novelty bias, might give non-experts the impression that they can be directly applied to a wide range of tasks.
% The reason is that the proactive creation process enhances understanding of view relations, thereby offering the potential to integrate workflows from creation to analysis. 
We need to note that it is still difficult to incorporate the proposed interactions in a professional context for data analysis.
% add discussion: actual analysis tasks and a respective workflow
%  expect a careful and better embedding of that information into the overall study discussion.
To achieve this integration,} it is crucial to ensure the consistency of interaction semantics at different stages and provide clear distinctions to convey the user's intentions. \zq{This requires accommodating} more types interaction commands. One potential way is to divide visualization views into more detailed interaction targets (Fig.~\ref{fig:DesignSpace}).% As shown in , where visualization views are subdivided into various interaction targets, 
Future IA systems could benefit from providing distinctions in terms of grasp areas or angles for manipulating composite views. 

\textbf{Facilitating Flexibile Data Analysis and Communication} through Interaction.
Our study recognizes that users prefer to actively create composite views because of the freedom to combine and separate views. Previously, composite views in immersive environments were meticulously designed by visualization experts based on user requirements, without providing users the ability to manually merge or separate different views~\cite{yang2020tilt, liu2020design}. Our work introduces a novel perspective for designers and developers of Immersive Analytics (IA) systems. However, our current approach is not yet directly applicable to user data analysis tasks.
% interactively composing visualizations to empower users, especially data analysts, to rapidly validate their ideas. For example, it can help data analysts quickly test and refine ideas in the early design stages of composite visualizations.
We advocate for future IA systems to provide users with a more adaptable and interactive approach to validate their ideas during the analysis of composite visualizations. 
% This could involve functionalities such as combining two visual views to assess their potential combinations. 
Such support requires the computation of underlying data semantics and data transformations, followed by the formulation of concrete design guidelines as constraints, informed by a multitude of use cases~\cite{deng2022revisiting}.

\textbf{Generative Capability of the Design Space.}
To develop the design space, our research begins by exploring existing cases of composite views~\cite{deng2022revisiting}. Therefore, for visual representation, our design space focuses on describing composite views based on existing design patterns~\cite{javed2012exploring}.
Regarding generative capabilities, our design space focuses on interactions and offers flexible combinations of basic 3D manipulations, which can be attached to different interaction targets to customize interactions.
Although we have not studied new visual representations of composite views, we believe this is a promising direction, as combining visualization views in space offers new opportunities.
The number of existing immersive composite visualizations available for analysis is still limited. Therefore, it is worthwhile to investigate further how to leverage spatial environments to combine multiple visualization views.
% We plan to adopt formative methods (e.g., elicitation studies) to explore the possibilities of generating novel immersive visualizations.}
% The spatial properties and 3D rendering capabilities in immersive environments offer significant potential for designing innovative composite visualizations~\cite{yang2020tilt, yang2018origin}. 

\textbf{Potential Usage Scenarios.}
Based on the study findings, we found that building an interactive experience for creating composite visualization can be well-suited for several scenarios. 
This active user participation has the potential to provide an efficient, engaging, and compelling user experience.
For example, in educational settings, we can guide users to actively create composite views, which may provide users with deeper insights into data relations compared to simply presenting them with pre-designed composite visualizations.
% Our study indicated that participants generate their own ideas and understandings when they interactively build composite visualizations. For example, in educational settings, we can guide users to create corresponding data visualizations interactively through building such interactive cases or systems. This approach allows users to better understand the insights within the data compared to simply presenting them with pre-designed composite visualizations.
Furthermore, in the context of visual literacy education, this natural and intuitive interactive experience can assist non-experts in interpreting unfamiliar composite visualizations. 
Future IA systems may provide step-by-step interactive building experiences, allowing users to gradually build composite views~\cite{lee2016vlat}. 
% This has the potential to promote and enhance people's visual literacy, or facilitate training in visualization-related abilities.
Moreover, designers or developers need to provide proper guidance for users to understand the semantic meaning conveyed by interactions and data relations.
% consider which information needs to be transparent, and which information can be hidden to reduce user cognitive load when designing interactive experiences. 
Insufficient guidance and introduction regarding data semantics or motivations of interactions may result in users solely manipulating elements without a clear understanding of data insights embedded in the interactive progress.
% Composite visualizations often increase the complexity of visual interpretation, requiring higher levels of visual literacy. The interactive building experience, starting from basic views that users can comprehend and gradually constructing composite visualizations through interaction~\cite{lee2016vlat}, has the potential to promote and enhance visual literacy. 
% Nevertheless, before providing interactive experiences, users need to have a deeper understanding of the data relationships. The design of IA systems needs to consider which information needs to be transparent and which does not during the construction of composite visualizations, depending on specific users and tasks.

% From the perspective of data communication and presentation, this interactive experience plays a crucial role in enhancing communication and storytelling with composite visualizations.
% Presenters can integrate the primitive views and demonstrate step-by-step how to construct the corresponding composite visualizations. Moreover, it encourages audience participation, empowering them to interactively contribute insights and effectively convey their perspectives.
In addition, this interactive experience may apply to collaborative work or design process. 
Imagine a scenario where numerous individuals leverage visualization in meetings. 
For example, in analyzing data from a vast social network depicted as a graph. Each participant can select nodes of interest, introduce events, or highlight character relationships they wish to emphasize. Participants can articulate their perspectives by seamlessly composing several visualization views into a single visualization and demonstrating them to other collaborators.
This dynamic process may effectively alleviate the working memory load, facilitating the organization and consolidation of ideas.
% Each person can select nodes of interest and insert their desired subviews. This dynamic process can reduce working memory load, making collaboration and meetings more efficient when organizing and consolidating ideas and thoughts.

\textbf{Limitations and Future Work.}
% Our work investigates the creation of composite visualizations through embodied interactions. 
% Our findings can provide new perspectives or serve as references for system designers and developers in the thriving field of immersive analysis.
There are several limitations in our study.
First, our research primarily revolves around the conceptualization and data relationships of composite visualizations, while overlooking the design of visual representations in immersive environments. 
This limitation stems from the absence of established design standards for composite visualizations in immersive environments. We will consider developing novel composite visualization cases to establish relevant design guidelines.
% This is primarily due to the lack of established visual design standards in immersive environments. Past principles of visual design can serve as a reference point~\cite{tufte2001visual}. In the future, it will be necessary to conduct research and summarize more cases of composite visualizations within immersive environments to derive design guidelines.
Additionally, our research mainly considers data connections between different visualization views, neglecting other potential relationships, such as temporal or hierarchical relationships~\cite{von2012visual, gotz2019visual}. Future work could combine other types of data relationships to design dynamic interactive experiences for composite visualizations.
Lastly, the design space represents an exploratory effort to enable users to actively create composite visualizations. We explored five representative types of composite visualizations, without delving into the composition of multiple composite views, which may involve more complex design scenarios. 
% We also plan to evaluate the design space utility by working with visualization experts.
Future work should continue to advance the development of relevant composite visualization designs tailored to concrete tasks or usage scenarios.
% Constrained by device computational capabilities and a lack of established design principles in immersive analytics~\cite{ens2021grand}, 
% we have not yet been able to design and implement an effective system allows users to arbitrarily author novel composite visualizations of any possible and unknown visual expressions. 

% consistency with the following interaction for data analysis to achieve fluid experience
% employ physical metaphor with visualization, make virtual vis as a physical object with affordance
% analogy with life exp.
% more customize from creating stage rather than analyzing stage...

%% file: sections/8_Conclusion.tex
\section{Conclusion}
We explore embodied interactions for creating composite visualizations in immersive environments. 
% We start by distilling data relationships of visualization views and mapping them to the spatial combinations of composite views. 
We formulate the composition of visualization based on the constraints of data relations and user interaction. 
Then, we develop a design space of embodied interactions for compositing visualizations, which considers interaction targets, direct manipulation, and interaction-induced view relations. 
Finally, we demonstrate the design space with representative cases. 
Through a user study that evaluates the usability and user experience with the interactions, we discuss the key insights for future immersive analytics systems.

%% file: sections/9_Ack.tex
\section{Acknowledgment}
We wish to thank all members of the Georgia Tech Visualization Lab for their feedback on our paper. We also wish to thank all our participants
for their time and our reviewers for their comments and feedback. 
This work is supported by the Research Grants Council of the Hong Kong Special Administrative Region under General Research Fund (GRF) with Grant No. 16207923.